\documentclass[aip,amsmath,amssymb,reprint]{revtex4-1}

\usepackage{graphicx}
\usepackage{dcolumn}
\usepackage{bm}
\usepackage[T1]{fontenc}
\usepackage{xcolor}
\usepackage{float}
\usepackage{pgfplotstable}
\usepackage{array}
\usepackage{booktabs}
\usepackage[export]{adjustbox}
\usepackage{hyperref}
\usepackage{soul}
\usepackage[version=4]{mhchem}
\usepackage{amssymb}
\usepackage{braket}
\usepackage{xspace}
\usepackage{comment}
\usepackage{newfloat} 
\DeclareFloatingEnvironment[
  fileext = los ,
  listname = {List of Schemes},
  name = SCHEME
]{scheme}

\newcommand{\ie}[0]{\textit{i.e.}, }
\newcommand{\eg}[0]{\textit{e.g.}, }
\newcommand{\cf}[0]{\textit{cf}.\ }
\newcommand{\via}[0]{\textit{via} }
\newcommand{\norm}[1]{\left\lVert#1\right\rVert}
\usepackage{scalefnt}

\makeatletter
\newsavebox\myboxA
\newsavebox\myboxB
\newlength\mylenA
\newcommand*\xoverline[2][0.75]{%
    \sbox{\myboxA}{$\m@th#2$}%
    \setbox\myboxB\null
    \ht\myboxB=\ht\myboxA%
    \dp\myboxB=\dp\myboxA%
    \wd\myboxB=#1\wd\myboxA
    \sbox\myboxB{$\m@th\overline{\copy\myboxB}$}
    \setlength\mylenA{\the\wd\myboxA}
    \addtolength\mylenA{-\the\wd\myboxB}%
    \ifdim\wd\myboxB<\wd\myboxA%
       \rlap{\hskip 0.5\mylenA\usebox\myboxB}{\usebox\myboxA}%
    \else
        \hskip -0.5\mylenA\rlap{\usebox\myboxA}{\hskip 0.5\mylenA\usebox\myboxB}%
    \fi}
\makeatother
\DeclareMathOperator{\Tr}{Tr}
\newcommand{\DefineAuthor}[2]{%
  \expandafter\newcommand\csname #1note\endcsname[1]{%
    \textbf{\textcolor{#2}{\textbf{#1:} ##1}}}%
  \expandafter\newcommand\csname #1\endcsname[1]{
    \textbf{\textcolor{#2}{##1}}}
  \expandafter\newcommand\csname #1cancel\endcsname[1]{%
    \textbf{\textcolor{#2}{\sout{##1}}}}%
  \expandafter\newcommand\csname #1change\endcsname[2]{%
    \textbf{\textcolor{#2}{\sout{##1} ##2}}}%
  \newenvironment{#1text}{\color{#2}}{\color{black}}
}
\definecolor{dartmouthgreen}{rgb}{0.05, 0.5, 0.06}
\definecolor{trinered}{rgb}{0.8, 0.0, 0.0}
\DefineAuthor{RAD}{red}
\DefineAuthor{BGE}{blue}
\DefineAuthor{ZMS}{magenta}
\DefineAuthor{TKQ}{orange}
\DefineAuthor{GEN}{dartmouthgreen}
\DefineAuthor{FIG}{dartmouthgreen}
\makeatletter
\let\@fnsymbol\@fnsymbol@latex
\@booleanfalse\altaffilletter@sw
\makeatother

\def\gobbledot.{} 

\begin{document} 

\title{CASE21: Uniting Non-Empirical and Semi-Empirical Density Functional Approximation Strategies using Constraint-Based Regularization}

\author{Zachary M. Sparrow}
\author{Brian G. Ernst}
\author{Trine K. Quady}
\author{Robert A. DiStasio Jr.} 
\email{distasio@cornell.edu}
\affiliation{Department of Chemistry and Chemical Biology, Cornell University, Ithaca, NY 14853 USA}

\date{\today}


\begin{abstract} 
In this work, we present a general framework that unites the two primary strategies for constructing density functional approximations (DFAs): non-empirical (NE) constraint satisfaction and semi-empirical (SE) data-driven optimization.
The proposed method employs B-splines---bell-shaped spline functions with compact support---to construct each inhomogeneity correction factor (ICF).
This choice offers several distinct advantages over a polynomial basis by enabling explicit enforcement of linear and non-linear constraints as well as ICF smoothness using Tikhonov regularization and penalized B-splines (P-splines).
As proof of concept, we use this approach to construct CASE21---a \textbf{C}onstrained \textbf{A}nd \textbf{S}moothed semi-\textbf{E}mpirical hybrid generalized gradient approximation that completely satisfies all but one constraint (and partially satisfies the remaining one) met by the PBE0 NE-DFA and exhibits enhanced performance across a diverse set of chemical properties.
As such, we argue that the paradigm presented herein maintains the physical rigor and transferability of NE-DFAs while leveraging high-quality quantum-mechanical data to improve performance.
\end{abstract}

\maketitle

Kohn-Sham density functional theory (KS-DFT) is the \textit{de facto} standard for electronic structure calculations in chemistry, physics, and materials science due to its favorable trade-off between accuracy and computational cost.~\cite{mardirossianThirtyYearsDensity2017}
While there now exist hundreds of density functional approximations (DFAs) of varying complexity across all rungs of Perdew's popular Jacob's ladder,~\cite{perdewJacobLadderDensity2001} most have been designed using either non-empirical (NE) or semi-empirical (SE) strategies.~\cite{beckeDensityfunctionalThermochemistrySystematic1997,perdewGeneralizedGradientApproximation1996,mardirossianThirtyYearsDensity2017}
NE strategies seek to construct DFAs by proposing simple ans{\"a}tze designed to satisfy well-defined physical constraints (\eg the uniform electron gas (UEG) limit,~\cite{kurthMolecularSolidstateTests1999} second-order gradient responses~\cite{wangSpinScalingElectrongas1991,maCorrelationEnergyElectron1968,geldartExchangeCorrelationEnergy1976,langrethTheoryNonuniformElectronic1980}).
Resulting NE-DFAs (\eg PBE,~\cite{perdewGeneralizedGradientApproximation1996} PBE0,~\cite{adamoReliableDensityFunctional1999} SCAN~\cite{sunStronglyConstrainedAppropriately2015}) tend to be more transferable across systems and favored in the physics and materials science communities.
SE strategies seek to construct DFAs by optimizing a physically motivated and flexible functional form to best reproduce high-quality reference quantum-mechanical data.
Resulting SE-DFAs (\eg B3LYP,~\cite{beckeDensityFunctionalThermochemistry1993} Minnesota functionals,~\cite{peveratiQuestUniversalDensity2014,zhaoExchangecorrelationFunctionalBroad2005,zhaoM06SuiteDensity2008} 
the B97 family~\cite{mardirossianThirtyYearsDensity2017,beckeDensityfunctionalThermochemistrySystematic1997,mardirossianOB97XV10parameterRangeseparated2014})
often perform quite well (typically exceeding NE-DFAs) on chemical systems/properties similar to the training data, and are popular for chemical applications.

When used independently, both of these DFA strategies have shortcomings.
For one, NE-DFA ans{\"a}tze are somewhat arbitrary, \ie there is some flexibility when constructing an NE-DFA that satisfies a given set of constraints.~\cite{sunStronglyConstrainedAppropriately2015,perdewPrescriptionDesignSelection2005}
Consequently, there is no guarantee that the chosen ansatz will perform best in practice.
In the same breath, the choice of constraints is also somewhat arbitrary/empirical, \eg the correct series expansion of the exchange-correlation (xc) energy is sometimes ignored as it often results in inaccurate DFAs for real systems.~\cite{yuPerspectiveKohnShamDensity2016}
On the other hand, striving for the best-performing functional using only an SE-DFA strategy often goes hand-in-hand with sacrificing exact physical constraints.~\cite{medvedevDensityFunctionalTheory2017,beckeDensityfunctionalThermochemistrySystematic1997,mardirossianOB97XV10parameterRangeseparated2014,peveratiQuestUniversalDensity2014}
Furthermore, some SE-DFAs suffer from non-physical ``bumps'' or ``wiggles'' in the inhomogeneity correction factor (ICF), which violate an \textit{implicit} smoothness constraint and can require significantly larger quadrature grids for accurate integration.~\cite{dasguptaStandardGridsHighprecision2017,mardirossianHowAccurateAre2016,wheelerIntegrationGrid2011,wheelerIntegrationGrid2019} 
Clearly, both paradigms provide useful information about the optimally performing DFA, but neither suffices on its own.

While several groups have advocated for combining these strategies,~\cite{yuPerspectiveKohnShamDensity2016,brownMCMLCombiningPhysical2021} constraint satisfaction during the data-driven optimization process has remained difficult to date. 
To address the smoothness problem in SE-DFAs, the BEEF~\cite{wellendorffDensityFunctionalsSurface2012,wellendorffMBEEFAccurateSemilocal2014,lundgaardMBEEFvdWRobustFitting2016} and Minnesota~\cite{vermaStatusChallengesDensity2020} functionals have adopted an explicit smoothness penalty in the regression procedure with reasonable success; the resulting ICFs are significantly smoother than previous generations, albeit not always completely devoid of spurious features.
Furthermore, the recent MCML approach~\cite{brownMCMLCombiningPhysical2021} has made efforts to combine NE-DFA and SE-DFA strategies by algebraically enforcing three linear constraints during the SE-DFA optimization process (an approach originally used in the M05 family~\cite{zhaoDesignDensityFunctionals2006}).
While successful in enforcing the targeted constraints, the polynomial basis used in MCML (and the vast majority of SE-DFAs to date) prevents explicit enforcement of non-linear constraints (such as inequalities), and makes satisfying new constraints non-trivial as each regression coefficient appears in every algebraic constraint.

In this work, we present a general framework that unites NE-DFA and SE-DFA strategies by enabling straightforward enforcement of both physical constraints and ICF smoothness while leveraging high-quality quantum-mechanical data.
The proposed DFA strategy uses B-splines, compact bell-shaped piece-wise functions,~\cite{eilersFlexibleSmoothingBsplines1996} to construct the ICF, which allows for a tunable trade-off between ICF smoothness and flexibility using penalized B-spline (P-spline) regularization,~\cite{eilersTwentyYearsPsplines2015} while still allowing for explicit enforcement of both linear and non-linear constraints \via generalized Tikhonov regularization.
As proof of concept, we use this framework to construct a hybrid generalized gradient approximation (GGA):  CASE21---\textbf{C}onstrained \textbf{A}nd \textbf{S}moothed semi-\textbf{E}mpirical 20\textbf{21}, which completely satisfies all but one constraint (and partially satisfies the remaining one) met by the PBE0 NE-DFA.
When compared to PBE0 (and the popular B3LYP SE-DFA), CASE21 attains higher accuracy across a diverse set of chemical properties without sacrificing transferability or requiring large numerical quadrature grids.
As such, we argue that the CASE paradigm presented herein maintains the physical rigor and transferability of NE-DFAs while leveraging high-quality quantum-mechanical data to remove the arbitrariness of ansatz selection.

\textit{Functional Form.} 
We write CASE21 as the sum of exchange and correlation contributions, 
\begin{align}
    E_{\rm xc}^{\rm CASE21} = \frac{3}{4} E_{\rm x}[\rho_\uparrow, \rho_\downarrow] + \frac{1}{4} E_{\rm xx} + E_{\rm c}[\rho, \zeta] ,
    \label{eq:xc_energy}
\end{align}
where the exchange contribution uses $25\%$ exact exchange ($E_{\rm xx}$), as generally recommended for global hybrid GGAs.~\cite{adamoReliableDensityFunctional1999,ernzerhofAssessmentPerdewBurke1999} 
The semi-local exchange is defined using the exchange spin scaling relationship:~\cite{oliverSpindensityGradientExpansion1979} 
\begin{align}
    E_{\rm x}[\rho_\uparrow,\rho_\downarrow] = \frac{1}{2} \big( E_{\rm x}[2\rho_\uparrow] + E_{\rm x}[2\rho_\downarrow] \big) ,
    \label{eq:ex_spin_scaling}
\end{align}
in which
\begin{align}
    E_{\rm x}[\rho_\sigma] = \int \rho_\sigma \, \epsilon_{\rm x}^{\rm LDA}(\rho_\sigma) F_{\rm x}(u_{\rm x,\sigma}) \, d\bm{r} ,
    \label{eq:exch}
\end{align}
$\rho_\sigma$ is the spin density (with spin $\sigma \in \{ \uparrow, \downarrow \}$), $\epsilon_{\rm x}^{\rm LDA}$ is the exchange energy density per particle within the local density approximation (LDA), and $F_{\rm x}(u_{\rm x,\sigma})$ is the yet to be determined CASE21 exchange ICF.
We employ $0 \le u_{\rm x, \sigma} = (\gamma_{\rm x} s_\sigma^2)/(1+\gamma_{\rm x} s_\sigma^2) < 1$ (as originally proposed by Becke~\cite{beckeDensityFunctionalCalculations1986}) as the finite-domain representation of the PBE dimensionless spin density gradient, $s_\sigma = |\nabla \rho_\sigma|/[\pi^{2/3}(2\rho_\sigma)^{4/3}]$.
Here, we note that the PBE exchange ICF can be written as a linear function of $u_{\rm x,\sigma}$ if $\gamma_{\rm x} = \mu/\kappa \approx 0.273022$ (where $\mu$ and $\kappa$ are the NE parameters in PBE), which we denote by $\xoverline{F}_{\rm x}(u_{\rm x,\sigma}) \equiv 1 + \kappa u_{\rm x, \sigma}$.
Hence, we argue that this is an appropriate choice for $\gamma_{\rm x}$ since the
UEG exchange limit,~\cite{kurthMolecularSolidstateTests1999} UEG linear response,~\cite{perdewGeneralizedGradientApproximation1996} and Lieb-Oxford bound~\cite{liebImprovedLowerBound1981} can still be straightforwardly enforced in this smooth limiting form (\textit{vide infra}).

We construct $E_{\rm c}[\rho,\zeta]$ by analogy to $E_{\rm x}[\rho_\sigma]$, namely,
\begin{align}
    E_{\rm c}[\rho, \zeta] = \int \rho \, \epsilon_{\rm c}^{\rm LDA}(\rho, \zeta) F_{\rm c}(u_{\rm c}) \, d\bm{r} ,
    \label{eq:corr}
\end{align}
in which $\epsilon_{\rm c}^{\rm LDA}(\rho, \zeta)$ is the PW92~\cite{perdewAccurateSimpleAnalytic1992} LDA correlation energy density per particle, $\rho = \rho_\uparrow + \rho_\downarrow$ is the total density, $\zeta = (\rho_\uparrow - \rho_\downarrow)/\rho$ is the relative spin polarization, and $F_{\rm c}(u_{\rm c})$ is the yet to be determined CASE21 correlation ICF.
As with exchange, we suggest a form for $u_c$ such that a linear ICF, \ie $\xoverline{F}_c(u_{\rm c}) \equiv 1 - u_{\rm c}$, would satisfy the
UEG correlation limit,~\cite{kurthMolecularSolidstateTests1999}
rapidly varying density limit,~\cite{perdewGeneralizedGradientApproximation1996}
and second-order gradient expansion for correlation.~\cite{wangSpinScalingElectrongas1991,maCorrelationEnergyElectron1968,geldartExchangeCorrelationEnergy1976,langrethTheoryNonuniformElectronic1980}
Namely, we propose $0 \le u_{\rm c} \equiv (-\phi^3 t^2)/(-\phi^3 t^2 + \gamma_{\rm c} \epsilon_{\rm c}^{\rm LDA}) < 1$, where $\phi = \frac{1}{2} \left[ (1+\zeta)^{2/3} + (1-\zeta)^{2/3} \right]$ is a spin scaling factor,~\cite{wangSpinScalingElectrongas1991} $\gamma_{\rm c} = 1/\beta \approx 14.986886$ (where $\beta$ is another NE parameter in PBE), and $t$ is a dimensionless spin-separated density gradient,
\begin{align}
    t \equiv \sqrt{a_0} \left( \frac{\pi}{3} \right)^{1/6} \frac{|\nabla \rho_\uparrow|+|\nabla \rho_\downarrow|}{4 \rho^{7/6} \phi} ,
    \label{eq:t_def}
\end{align}
which reduces to the PBE dimensionless density gradient ($t^{\rm PBE}$, which has $|\nabla \rho|$ instead of $|\nabla \rho_\uparrow|+|\nabla \rho_\downarrow|$ in the numerator) when $|\nabla \zeta| = 0$ (which was assumed during the construction of PBE correlation, and is a relationship that allows DFAs based on $t$ to satisfy PBE correlation constraints).
We note in passing that the use of $t^{\rm PBE}$ yields qualitatively similar results to $t$ (which might be expected, given that $t$ and $t^{\rm PBE}$ are equivalent for closed-shell systems), although $t$ slightly outperforms $t^{\rm PBE}$ quantitatively.
Importantly, $u_{\rm c}$ increases monotonically with $t$, suggesting a one-to-one mapping between $t$ and $u_{\rm c}$ for a given $\epsilon_{\rm c}^{\rm LDA}$; hence, $u_{\rm c}$ is an appropriate finite-domain transformation of $t$.
While Eq.~\eqref{eq:corr} with this definition of $u_{\rm c}$ does not fully satisfy uniform scaling to the high-density limit for correlation,~\cite{levyAsymptoticCoordinateScaling1989} it does completely cancel the $\epsilon_{\rm c}^{\rm LDA}$ logarithmic singularity~\cite{uniscalenote} and allows for satisfaction of all other PBE correlation constraints.
However, such partial satisfaction of this constraint is not a restriction of the presented method---in principle, an (albeit more complex) functional form that completely satisfies all PBE correlation constraints could have also been used.

\begin{figure}[t]
    \centering
    \includegraphics[width=1.0\linewidth]{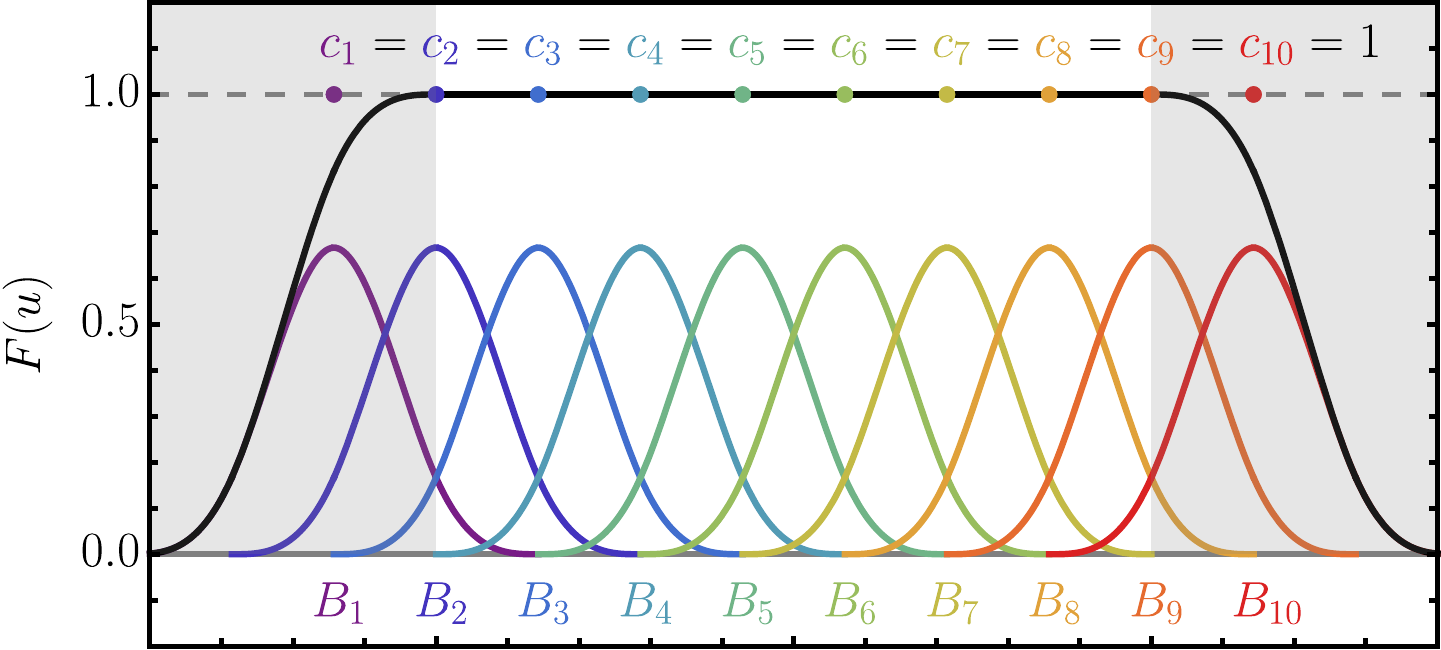} \\[0.5em]
    \includegraphics[width=1.0\linewidth]{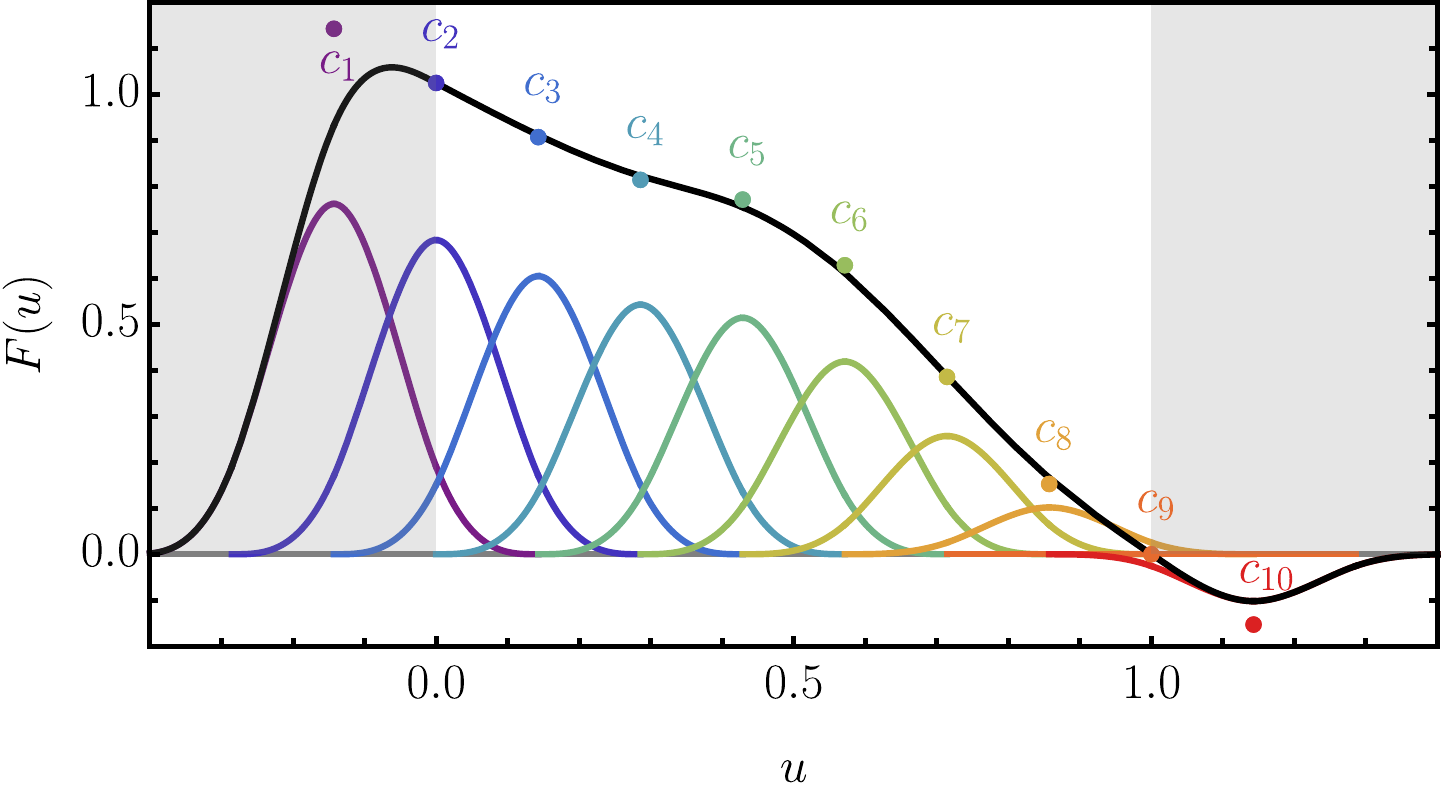}
    \caption{
    (\textit{Top}) B-spline basis functions ($\{ B_{i} \}_{i=1,10}$, rainbow) used to represent the exchange and correlation ICFs in this work. When all expansion coefficients are set to unity, the resulting B-spline curve ($F(u) = \sum_i c_i B_i(u)$, black) is uniform in $0 \leq u \leq 1$ and recovers the LSDA/LDA limit.  
    (\textit{Bottom}) B-spline curve with non-uniform coefficients. Note again how the coefficients closely align with the curve for $0 \leq u \leq 1$.
    }
    \label{fig:bspline}
\end{figure}

We write the CASE21 exchange and correlation ICFs as linear combinations of $N_{\rm sp}$ compact piece-wise bell-shaped cubic ($k = 3$) uniform B-spline basis functions ($\{ B_i \}$),~\cite{eilersFlexibleSmoothingBsplines1996} 
\begin{align}
    \begin{split}
        F_{\rm x}(u_{\rm x,\sigma}) &= \sum_{i}^{N_{\rm sp}} c_{{\rm x},i} B_i(u_{\rm x, \sigma}) = \bm{c}_{\rm x} \cdot \bm{B}_{\rm x,\sigma} \\
        F_{\rm c}(u_{\rm c}) &= \sum_i^{N_{\rm sp}} c_{\rm c,i} B_i(u_{\rm c}) = \bm{c}_{\rm c} \cdot \bm{B}_{\rm c} ,
        \label{eq:icf}
    \end{split}
\end{align}
which is equivalent to constructing each ICF using a cubic spline~\cite{prautzschBezierBSplineTechniques2002} (see \textit{Supporting Information} (SI) for more details).
With the choice of knot vector employed herein,~\cite{eilersFlexibleSmoothingBsplines1996,eilersTwentyYearsPsplines2015} the $B_i(u_{\rm x, \sigma})$ and $B_i(u_{\rm c})$ are uniformly spaced with all points in $0 \le u_{\rm x, \sigma} \le 1$ and $0 \le u_{\rm c} \le 1$ supported by three non-zero B-splines.
As depicted in Fig.~\ref{fig:bspline}(a), setting $\bm{c}_{\rm x} = \bm{1} = \bm{c}_{\rm c}$ in Eq.~\eqref{eq:icf} results in $F_{\rm x}(u_{\rm x,\sigma}) = 1 = F_{\rm c}(u_{\rm c})$; in this limit, CASE21 exchange and correlation reduce to LSDA exchange and LDA correlation, respectively. 
%
%
%
%

Having defined the CASE21 functional form, we now discuss a general framework that unites NE-DFA and SE-DFA strategies.
Namely, we determine $\bm{c} = (\bm{c}_{\rm x}, \bm{c}_{\rm c})$ using generalized Tikhonov regularization,~\cite{hansenRankDeficientDiscreteIllPosed1998} \ie by minimizing the following loss function:
\begin{align}
    \mathcal{L} = \norm{\bm{X}\bm{c} - \bm{y}}_{\bm{W}}^2 + \lambda \norm{\bm{c}}_{\bm{A}}^2 + \eta \sum_i\norm{\bm{c} - \bm{c}_0}_{\bm{Q}_i}^2 ,
    \label{eq:loss}
\end{align}
where $\norm{\bm{v}}_{\bm{M}}^2 = \bm{v}^\top \bm{M} \bm{v}$ is the matrix norm of the vector $\bm{v}$ using the matrix $\bm{M}$, the sum is over the enforced constraints, and all other quantities will be defined below. 
Hence, the key to determining $\bm{c}$ lies in appropriate matrix norm choices in each term in $\mathcal{L}$: goodness of fit, regularization/smoothness, and constraint satisfaction.

\textit{Goodness of Fit.}
In the goodness of fit term, we construct the design matrix $\bm{X}$ by first noting that substitution of  Eq.~\eqref{eq:icf} into Eqs.~\eqref{eq:exch}--\eqref{eq:corr} (with fixed orbitals) casts $E_{\rm x}[\rho_\sigma]$ and $E_{\rm c}[\rho, \zeta]$ into linear forms in $\bm{c}_{\rm x}$ and $\bm{c}_{\rm c}$:
\begin{align}
    \begin{split}
        E_{\rm x}[\rho_\sigma] &= \sum_i^{N_{\rm sp}} c_{{\rm x},i} \int \rho_\sigma \, \epsilon_{\rm x}^{\rm LDA}(\rho_\sigma) B_i(u_{\rm x, \sigma}) \,d\bm{r} \equiv \bm{c}_{\rm x} \cdot \bm{\xi}_{{\rm x},\sigma} \\
        E_{\rm c}[\rho, \zeta] &= \sum_i^{N_{\rm sp}} c_{{\rm c},i} \int \rho \, \epsilon_{\rm c}^{\rm LDA}(\rho, \zeta) B_i(u_{\rm c}) \, d\bm{r} \equiv \bm{c}_{\rm c} \cdot \bm{\xi}_{{\rm c}} .
        \label{eq:xi_def}
    \end{split}
\end{align}
Linear combinations of $\bm{\xi}_{\rm x, \sigma}$ and $\bm{\xi}_{\rm c}$ can then be used to construct \textit{semi-local} xc contributions to energy differences $\Delta E_{\rm xc}$ (\eg atomization energies, reaction energies, barrier heights) in a form amenable to linear regression using reference quantum-mechanical data.
Defining $\bm{\xi} \equiv (\bm{\xi}_{\rm x}, \bm{\xi}_{\rm c})$, with $\bm{\xi}_{\rm x}$ obtained after applying Eqs.~\eqref{eq:xc_energy}--\eqref{eq:ex_spin_scaling} to $\bm{\xi}_{\rm x, \uparrow}$ and $\bm{\xi}_{\rm x, \downarrow}$ in Eq.~\eqref{eq:xi_def}, allows us to write: 
\begin{align}
    \Delta E_{\rm xc} &= \sum_{j} \nu_{j} \left( \bm{c} \cdot \bm{\xi}_{j} \right) = \bm{c} \cdot \sum_{j} \nu_{j} \, \bm{\xi}_{j} \equiv \bm{c} \cdot \bm{x} ,
    \label{eq:x_def}
\end{align}
in which $\nu_j$ is the stoichiometric coefficient for the $j$-th component in $\Delta E_{\rm xc}$ (\ie the energy of a molecule or atom) and $\bm{x}$ is a single row of $\bm{X}$.
$\bm{y}$ is the corresponding vector of reference energy differences $\Delta E_{\rm xc}^{\rm ref}$, and our choice for $\bm{W}$ (a square diagonal matrix of weights $w_i \equiv 1/\Delta E_{{\rm xc},i}^{\rm ref}$) is motivated by the fact that the $\bm{c}$ minimizing the goodness of fit term only (\ie weighted least squares) is the best linear unbiased estimator (under some common assumptions) if the $w_i$ are inversely proportional to the variance in each measurement.~\cite{strutzDataFittingUncertainty2016}
Since $E_{\rm xc}$ is the only inexact term in KS-DFT, both bias- and variance-type DFA errors should scale linearly with $E_{\rm xc}$,~\cite{NENCI_part_II,NECI_2021} making this a natural choice for $\bm{W}$.
Here, we argue that the piece-wise nature of a B-spline curve offers more flexibility than the low-order polynomial expansions often used to represent SE-DFA ICFs (\eg the B97 family\cite{mardirossianThirtyYearsDensity2017,beckeDensityfunctionalThermochemistrySystematic1997,mardirossianOB97XV10parameterRangeseparated2014}); with the ability to conform to more subtle shapes, a B-spline ICF should be able to better leverage the reference data.

\textit{Regularization/Smoothness.} 
For the second term in $\mathcal{L}$, we note that B-splines can be regularized by explicitly penalizing deviations from smoothness (\ie ICF ``wiggles'') using P-splines, a regularization technique suggested by Eilers and Marx~\cite{eilersFlexibleSmoothingBsplines1996,eilersTwentyYearsPsplines2015} based on the observation that B-spline coefficients closely resemble the B-spline curve (see Fig.~\ref{fig:bspline}(b)). 
As such, smoothness can be explicitly enforced \via a finite-difference penalty on $\bm{c}$; in this work, we interpret non-smoothness as non-linearity in the ICF, and construct $\bm{A}$ from the second-derivative finite-difference matrix (see SI).
$\lambda$ is a hyperparameter that governs the relative importance of the regularization/smoothness and goodness of fit contributions to $\mathcal{L}$, and interpolates (assuming $\eta \ggg 1$, \textit{vide infra}) between linear ICFs (\ie $\xoverline{F}_{\rm x}(u_{\rm x, \sigma})$ and $\xoverline{F}_{\rm c}(u_{\rm c})$) that are completely constraint-driven (as $\lambda \rightarrow \infty$) and wiggly ICFs that are data-driven to the \textit{maximal} amount possible in this framework (as $\lambda \rightarrow 0$).
As such, any non-linearity in the final optimized CASE21 ICFs can be attributed to the data.
Here, we note that alternative interpretations of smoothness would result in penalizing other derivatives (\eg $F'''(u)$).
Separately penalizing the exchange and correlation ICFs (\ie using two $\lambda$-hyperparameters) is also possible if the regularization/smoothness contributions to $\mathcal{L}$ from $F_{\rm x}(u_{\rm x, \sigma})$ and $F_{\rm c}(u_{\rm c})$ strongly differ.
In this work, we found that P-spline regularization (which is uniquely enabled by the choice of a B-spline basis) yields ICFs devoid of any spurious ``wiggles'' \via single-$\lambda$ penalization of $F''(u)$ (\textit{vide infra}).
In contrast, an excessively large penalty (which results in decreased performance) is usually required to remove all non-physical ``bumps'' or ``wiggles'' in polynomial ICFs regularized \via Tikhonov (or ridge) regression.~\cite{wellendorffDensityFunctionalsSurface2012,s.yuNonseparableExchangeCorrelation2015}
Furthermore, although such polynomial-based smoothness penalties are somewhat effective in reducing DFA grid dependence,~\cite{vermaStatusChallengesDensity2020,mardirossianHowAccurateAre2016} these approaches have been largely ineffective when enforced alongside constraints.~\cite{wellendorffDensityFunctionalsSurface2012,petzoldConstructionNewElectronic2012}
On the other hand, we find no issues when simultaneously enforcing ICF smoothness as well as numerous linear and non-linear constraints.

\textit{Constraint Satisfaction.} 
During CASE21 construction, we fully enforce the following $10$ constraints: exchange spin scaling,~\cite{oliverSpindensityGradientExpansion1979} uniform density scaling for exchange,~\cite{levyHellmannFeynmanVirialScaling1985} UEG exchange limit,~\cite{kurthMolecularSolidstateTests1999} UEG linear response,~\cite{perdewGeneralizedGradientApproximation1996} Lieb-Oxford bound,~\cite{liebImprovedLowerBound1981} exchange energy negativity, UEG correlation limit,~\cite{kurthMolecularSolidstateTests1999} second-order gradient expansion for correlation,~\cite{wangSpinScalingElectrongas1991,maCorrelationEnergyElectron1968,geldartExchangeCorrelationEnergy1976,langrethTheoryNonuniformElectronic1980} rapidly varying density limit for correlation,~\cite{perdewGeneralizedGradientApproximation1996} and correlation energy non-positivity.~\cite{perdewGeneralizedGradientApproximation1996} 
We also partially enforce uniform scaling to the high-density limit for correlation~\cite{levyAsymptoticCoordinateScaling1989} (\textit{vide supra}).
In the constraint satisfaction term in $\mathcal{L}$, the $\{ \bm{Q_i} \}$ are chosen to measure constraint-specific deviations of $\bm{c}$ from $\bm{c}_{0}$, the coefficients corresponding to $\xoverline{F}_{\rm x}(u_{\rm x, \sigma})$ and $\xoverline{F}_{\rm c}(u_{\rm c})$.
Each $\bm{Q}_i$ corresponds to a constraint on $F(u)$ or $F'(u)$, and is constructed such that any constraint-satisfying $\bm{c}$ yields $\norm{\bm{c}-\bm{c}_0}_{\bm{Q}_i}^2 = 0$ (see SI for details on $\bm{Q}_i$ construction).
$\eta$ is a hyperparameter that governs the relative importance of the constraint satisfaction contribution to $\mathcal{L}$, and was chosen to be large enough ($\eta = 10^8$) for strict constraint satisfaction, but small enough to avoid conditioning issues.
Since each B-spline has compact support, each $\bm{Q}_i$ only enforces the constraint on a small subset of $\bm{c}$ (\eg the $\bm{c}$ corresponding to non-zero B-splines at the $u=0$ limit); in contrast, each constraint would generally involve every parameter in a polynomial-based ICF (\eg MCML~\cite{brownMCMLCombiningPhysical2021}).
Another important consequence of this local support is that the B-spline curve will lie within the range of $\bm{c}$ (\cf Fig.~\ref{fig:bspline}(b)).
Hence, inequality constraints can be enforced \via an iterative update to the corresponding $\bm{Q}_i$ using the shape constraint algorithm (SCA) of Bollaerts \textit{et al.},~\cite{bollaertsSimpleMultiplePsplines2006} which fixes all inequality-violating $c_i$ to the constraint boundary.
In contrast, there is no straightforward way to explicitly apply inequality constraints on a polynomial-based ICF as each basis function is unique and has global support.
This highlights another benefit provided by a B-spline basis in the construction of smooth and constraint-satisfying SE-DFAs.

\textit{Training Procedure.} 
Our self-consistent training procedure (Scheme~\ref{sch:procedure}, see \textit{Computational Methods} for more details) leverages three distinct data sets (see SI): training ($\bm{X}_{\rm train}, \bm{y}_{\rm train}$), validation ($\bm{X}_{\rm val}, \bm{y}_{\rm val}$), and testing ($\bm{X}_{\rm test}, \bm{y}_{\rm test}$).
In a given iteration, the training set (a single database of heavy atom transfer reaction energies, HAT707~\cite{kartonW411HighconfidenceBenchmark2011,mardirossianThirtyYearsDensity2017}) is used to initially determine $\bm{c}$ by minimizing $\mathcal{L}$ (in conjunction with the SCA for satisfying inequality constraints) for a range of $\lambda$ and a given set of orbitals $\{ \psi_i \}$ (with initial $\{ \psi_{i}^{0} \}$ generated using $\xoverline{F}_{\rm x}(u_{\rm x, \sigma})$ and $\xoverline{F}_{\rm c}(u_{\rm c})$). 
With $\bm{c}(\lambda)$, a weighted-root-mean-square error, 
\begin{align}
    {\rm wRMSE}(\lambda) = \sqrt{{\rm diag}(\bm{W})\cdot\bm{r}(\lambda)^2/\Tr{(\bm{W}})},
    \label{eq:wrmse}
\end{align}
in which $\bm{r}(\lambda)=\bm{X}_{\rm val}\bm{c}(\lambda) -\bm{y}_{\rm val}$ is the error vector and $\bm{r}(\lambda)^2$ is the element-wise square of $\bm{r}(\lambda)$, is computed on the validation set (which contains absolute energies of H--O from AE18~\cite{chakravortyGroundstateCorrelationEnergies1993,mardirossianThirtyYearsDensity2017} and all atomization energies in TAE203~\cite{kartonW417DiverseHighconfidence2017,morganteACCDBCollectionChemistry2019}).
Using $\lambda^* = {\rm argmin}_\lambda {\rm wRMSE}(\lambda)$, $\bm{c}^*$ is determined by re-optimizing $\mathcal{L}$ (in conjunction with the SCA) over the training and validation sets.
New $\{ \psi_i \}$ are then generated using $\bm{c}^*$, and the entire cycle is repeated until $\bm{c}^*$ is stationary.
At this point, the testing set (which contains a significantly more diverse range of chemical properties than the training and validation sets, \textit{vide infra}) is used to assess the performance and transferability of the self-consistent DFA.
%
%
\begin{scheme}[t]
    \centering
    \caption{
    Self-consistent DFA training procedure. 
    }
    \vspace{1em}
    \includegraphics[width=1.0\linewidth]{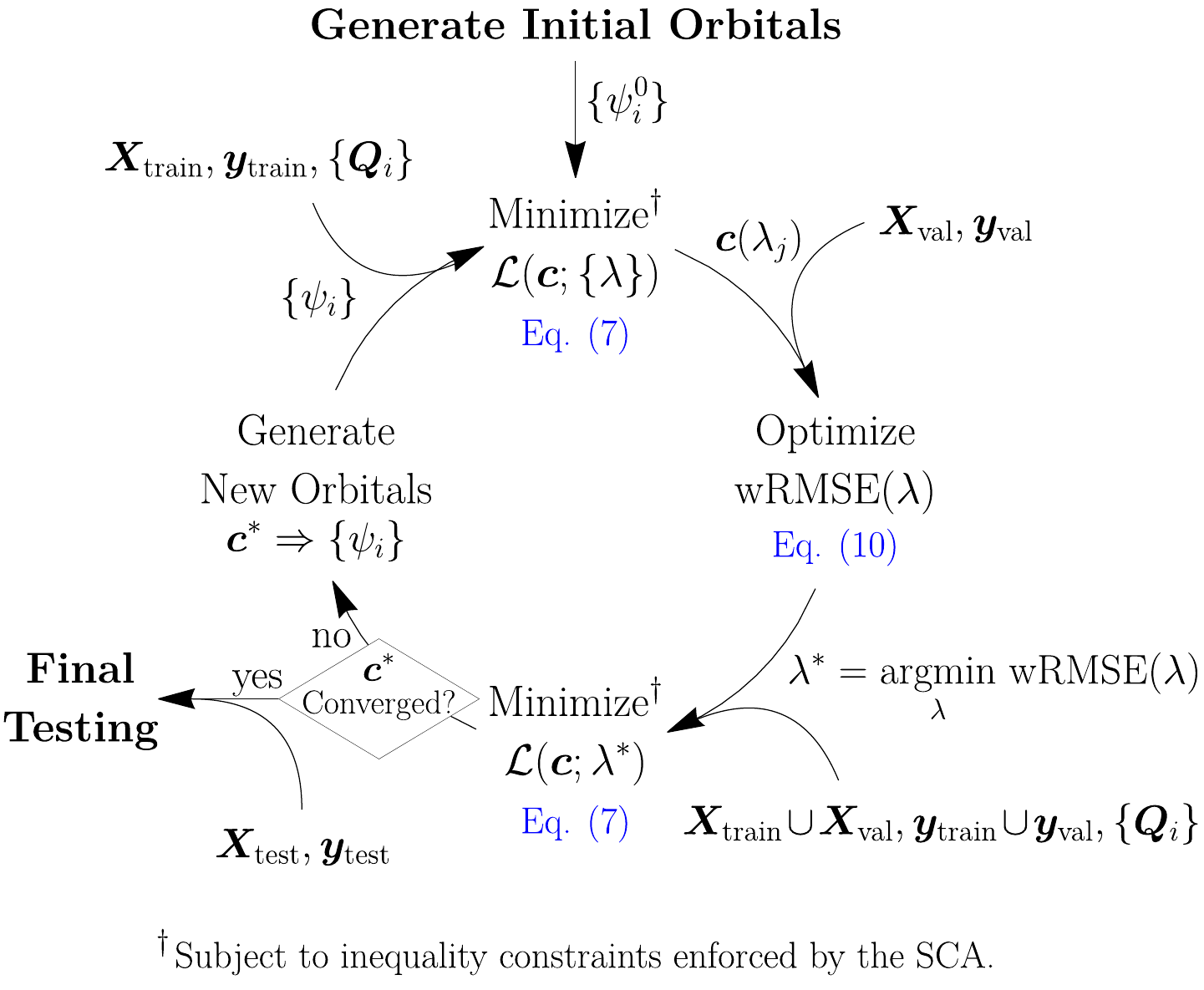} 
    \label{sch:procedure}
\end{scheme} 
%
%

\begin{figure}[t]
    \centering
    \includegraphics[width=1.0\linewidth]{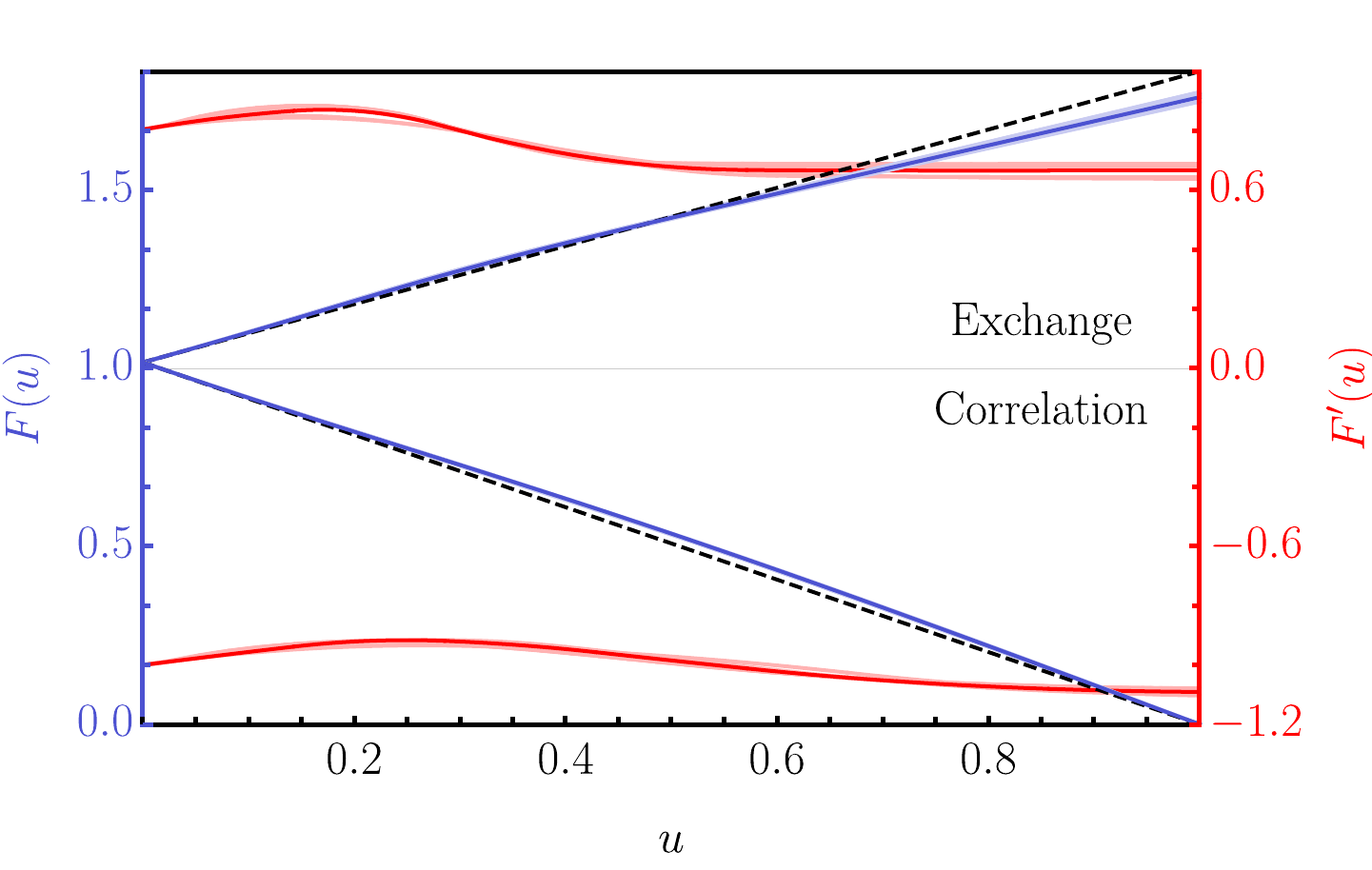} \\
    \includegraphics[width=1.0\linewidth]{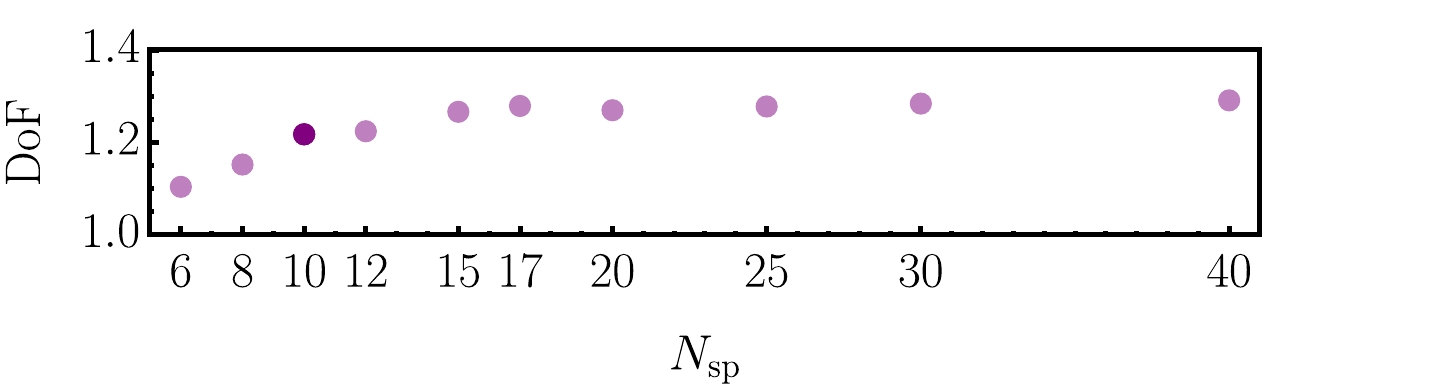} 
    \caption{
    (\textit{Top}) Exchange and correlation ICFs (blue) and first derivatives (red) for select $N_{\rm sp} \in [6,40]$.
    Highlighted curves (dark blue and dark red) correspond to the self-consistently optimized CASE21 DFA (with $N_{\rm sp} = 10$).
    Dashed lines represent the parameter-free linear ICFs ($\xoverline{F}_{\rm x}(u_{\rm x, \sigma})$ and $\xoverline{F}_{\rm c}(u_{\rm c})$) designed to satisfy the same constraints as CASE21.
    (\textit{Bottom}) Effective degrees of freedom (DoF) for select $N_{\rm sp} \in [6,40]$, with the dark purple point corresponding to CASE21.
    }
    \label{fig:icfs}
\end{figure}

Preliminary fits using $\{ \psi_{i}^{0} \}$ suggested that the \textit{effective} degrees of freedom~\cite{yeMeasuringCorrectingEffects1998} (DoF, see SI for derivation) change slowly starting around $N_{\rm sp} = 10$, and the performance of the corresponding (non-self-consistently optimized) DFA was representative of those with $N_{\rm sp} > 10$.
Hence, we used $N_{\rm sp} = 10$ in Scheme~\ref{sch:procedure} to generate the self-consistently optimized CASE21 DFA (six iterations; convergence criterion: $|\Delta\bm{c}| < 10^{-5}$; see SI for $\bm{c}^*$).
Even with a finite $\eta$, CASE21 nearly exactly satisfies all enforced constraints, \ie $F_{\rm x}(0)$, $F_{\rm c}(0)$, $F_{\rm x}'(0)$, and $F_{\rm c}'(0)$ differ from their corresponding exact values by ${\sim}10^{-5}$, $F_{\rm c}(1)$ differs by ${\sim}10^{-6}$, and all other constraints are exactly satisfied.
We therefore conclude that the proposed CASE framework (\ie Tikhonov regularization in conjunction with P-splines) successfully enforced all constraints without sacrificing smoothness, which still remains a challenge for other DFA training procedures.~\cite{wellendorffDensityFunctionalsSurface2012,petzoldConstructionNewElectronic2012}
To confirm that CASE21 remains representative of DFAs trained with other $N_{\rm sp}$ values, we (non-self-consistently) optimized $\bm{c}$ for select $N_{\rm sp} \in [6,40]$ using the CASE21 $\{ \psi_i \}$. 
As depicted in Fig.~\ref{fig:icfs}, the resulting ICFs and their first derivatives were all smooth and very similar (particularly for $N_{\rm sp} \geq 10$), thereby providing an \textit{a posteriori} justification for our choice of $N_{\rm sp} = 10$ for CASE21. 
From this plot, one can also see that the CASE21 ICFs (${\rm DoF} = 1.22$) subtly deviate from linearity in ways that simply cannot be obtained using low-order polynomial expansions.
%
%
%
%

%
%
\begin{figure}
    \centering
    \includegraphics[width=1.0\linewidth]{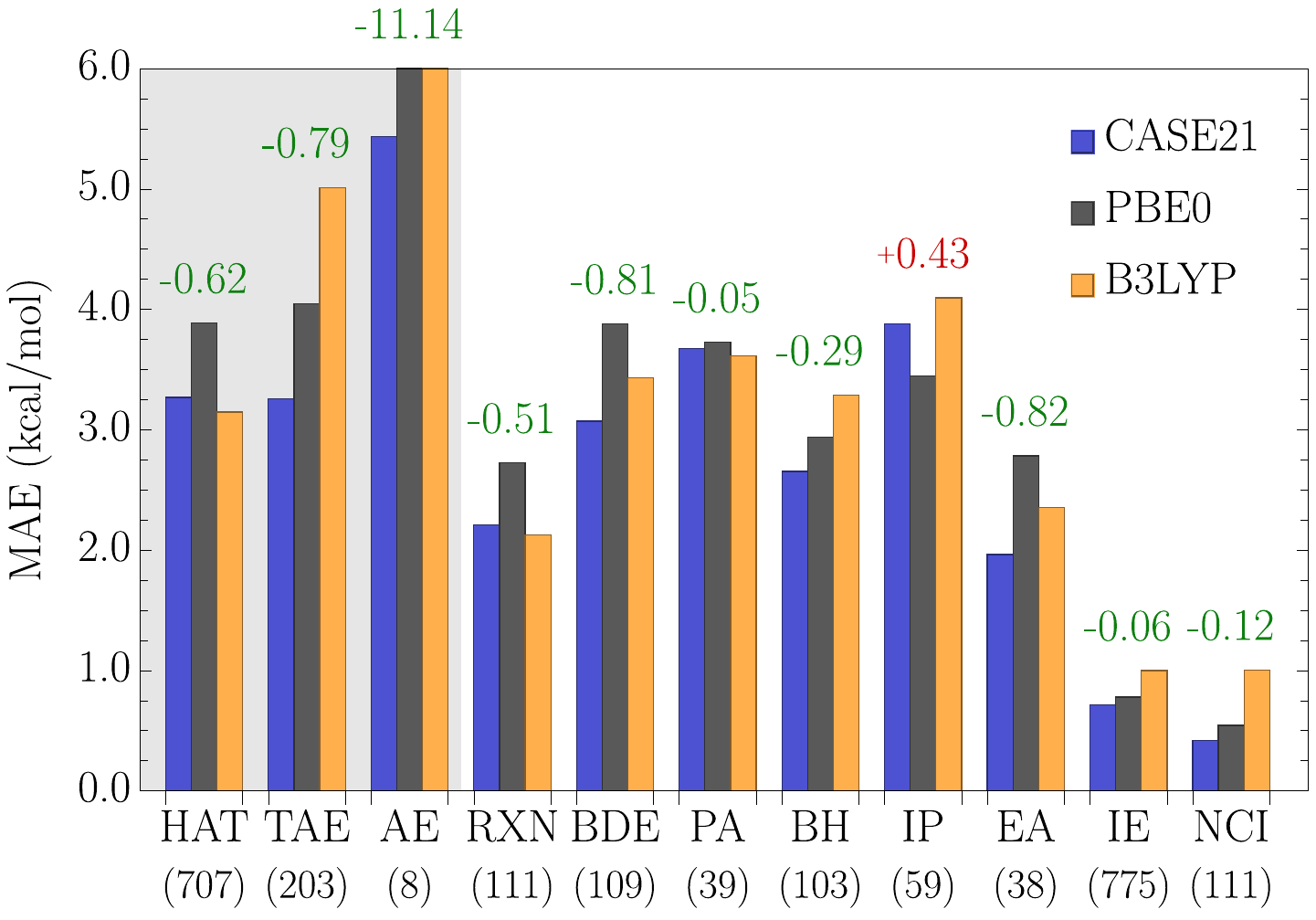} 
    \caption{
    Mean absolute errors of CASE21 (blue), PBE0 (gray), and B3LYP (orange) in the training/validation (shaded region) and testing (white region) sets.
    Bar labels indicate the relative performance of CASE21 and PBE0 (green: ${\rm MAE}^{\rm CASE21} < {\rm MAE}^{\rm PBE0}$; red: ${\rm MAE}^{\rm CASE21} > {\rm MAE}^{\rm PBE0}$).
    Properties (number of data points) include: 
    HAT---heavy atom transfer reaction energies,~\cite{kartonW411HighconfidenceBenchmark2011,mardirossianThirtyYearsDensity2017} 
    TAE---total atomization energies,~\cite{kartonW417DiverseHighconfidence2017,morganteACCDBCollectionChemistry2019} 
    AE---absolute energies,~\cite{chakravortyGroundstateCorrelationEnergies1993,mardirossianThirtyYearsDensity2017} 
    RXN---reaction energies,~\cite{goerigkGeneralDatabaseMain2010,zhaoBenchmarkDatabaseBarrier2005,zhaoDevelopmentAssessmentNew2004,mardirossianThirtyYearsDensity2017,goerigkLookDensityFunctional2017,curtissAssessmentGaussian2Density1997,morganteACCDBCollectionChemistry2019,neeseAssessmentOrbitalOptimizedSpinComponent2009,kartonW411HighconfidenceBenchmark2011,kartonAssessmentTheoreticalProcedures2012}
    BDE---bond dissociation energies,~\cite{yuComponentsBondEnergy2015,kartonW411HighconfidenceBenchmark2011,mardirossianThirtyYearsDensity2017,goerigkLookDensityFunctional2017,morganteACCDBCollectionChemistry2019} 
    PA---proton affinities,~\cite{goerigkLookDensityFunctional2017,parthibanAssessmentW1W22001,zhaoAssessmentDensityFunctionals2006,morganteACCDBCollectionChemistry2019}
    BH---barrier heights,~\cite{zhaoDevelopmentAssessmentNew2004,zhaoBenchmarkDatabaseBarrier2005,mardirossianThirtyYearsDensity2017,goerigkLookDensityFunctional2017,morganteACCDBCollectionChemistry2019}
    IP---ionization potentials,~\cite{curtissGaussianTheoryMolecular1991,lynchEffectivenessDiffuseBasis2003,goerigkGeneralDatabaseMain2010,goerigkLookDensityFunctional2017,morganteACCDBCollectionChemistry2019,mardirossianThirtyYearsDensity2017}
    EA---electron affinities,~\cite{lynchEffectivenessDiffuseBasis2003,curtissGaussianTheoryMolecular1991,goerigkGeneralDatabaseMain2010,mardirossianThirtyYearsDensity2017} 
    IE---isomerization energies,~\cite{gruzmanPerformanceInitioDensity2009,wilkeConformersGaseousCysteine2009,yuAssessmentTheoreticalProcedures2015,laoAccurateEfficientQuantum2015,grimmeHowComputeIsomerization2007,peveratiQuestUniversalDensity2014,luoValidationElectronicStructure2011,martinWhatCanWe2013,zhaoAssessmentDensityFunctionals2006,zhaoExchangecorrelationFunctionalBroad2005,zhaoNewLocalDensity2006,csonkaEvaluationDensityFunctionals2009,mardirossianPerformanceDensityFunctionals2013,kesharwaniBenchmarkInitioConformational2016,kartonW411HighconfidenceBenchmark2011,mardirossianThirtyYearsDensity2017,goerigkLookDensityFunctional2017,goerigkGeneralDatabaseMain2010,morganteACCDBCollectionChemistry2019} and 
    NCI---non-covalent interaction energies.~\cite{rezacDescribingNoncovalentInteractions2013,rezacAdvancedCorrectionsHydrogen2012,boeseAssessmentCoupledCluster2013,boeseBasisSetLimit2015,boeseDensityFunctionalTheory2015,smithBasisSetConvergence2014,kozuchHalogenBondsBenchmarks2013,mardirossianThirtyYearsDensity2017}
    }
    \label{fig:errors}
\end{figure}
%
%

\textit{Final Testing.} 
The performance of CASE21 across a diverse set of chemical properties is compared to that of the PBE0 and B3LYP hybrid DFAs in Fig.~\ref{fig:errors}.
CASE21 outperforms the PBE0 NE-DFA on $10/11$ properties, with improvements as large as $0.81$~kcal/mol and $0.82$~kcal/mol for bond dissociation energies and electron affinities, respectively.
In the testing set, CASE21 improves upon PBE0 in $7/8$ properties by an average of $0.38$~kcal/mol.
On the other hand, PBE0 only outperforms CASE21 for ionization potentials.
CASE21 also outperforms B3LYP (a popular SE-DFA for chemical applications) on $8/11$ properties; in the testing set, CASE21 improves upon B3LYP in $6/8$ properties by an average of $0.41$~kcal/mol (while B3LYP only offers a marginal ${\sim}0.07$~kcal/mol improvement on the remaining $2/8$).
We therefore conclude that CASE21 preserves the physical rigor and transferability of the PBE0 NE-DFA while still outperforming the B3LYP SE-DFA.
Although the CASE21 ICFs are clearly smooth (\cf Fig.~\ref{fig:icfs}), we also investigated the grid dependence of this DFA for completeness.
Since Lebedev-Treutler grids~\cite{treutlerEfficientMolecularNumerical1995} with $50$ radial and $194$ angular grid points (\ie $(50, 194)$) are typically large enough to obtain accurate energetics with standard hybrid GGAs (such as PBE0),~\cite{dasguptaStandardGridsHighprecision2017} we compared the performance of CASE21 using this grid to the larger grids employed during the training procedure (see \textit{Computational Methods}).
Using all points in the training, validation, and testing data sets ($N = 2{,}263$), we find nearly identical mean absolute deviations of $1.84 \times10^{-2}$~kcal/mol for CASE21 and $1.83 \times 10^{-2}$~kcal/mol for PBE0, thereby indicating that CASE21 does not require larger quadrature grids than PBE0 for accurate integration.

In this work, we presented the CASE (\textbf{C}onstrained \textbf{A}nd \textbf{S}moothed semi-\textbf{E}mpirical) framework for uniting NE-DFA and SE-DFA construction paradigms.
By employing a B-spline representation for the ICFs, this approach has several distinct advantages over the historical choice of a polynomial basis, namely, explicit enforcement of linear and non-linear constraints (using Tikhonov regularization) as well as explicit penalization of non-physical ICF ``bumps'' or ``wiggles'' (using P-splines).
As proof of concept, we used this approach to construct CASE21, a hybrid GGA that completely satisfies all but one constraint (and partially satisfies the remaining one) met by the PBE0 NE-DFA.
Despite being trained on only a handful of properties, CASE21 outperforms PBE0 and B3LYP (arguably the most popular SE-DFA for chemical applications) across a diverse set of chemical properties.
As such, we argue that the CASE framework can be used to design next-generation DFAs that maintain the physical rigor and transferability of NE-DFAs while leveraging benchmark quantum-mechanical data to remove the arbitrariness of ansatz selection and improve performance.
Future work will extend this approach to more sophisticated DFAs (\eg meta-GGAs, range-separated hybrids) as well as explore the use of B-splines in constructing robust features for machine-learning chemical properties.

\section*{Computational Methods}

All electronic structure calculations were performed using in-house versions of \texttt{Psi4 (v1.3.2)}~\cite{parrishPsi4OpenSourceElectronic2017} and \texttt{LibXC (v4.3.4)}~\cite{lehtolaRecentDevelopmentsLibxc2018a} modified with a self-consistent implementation of the CASE21 DFA (including functional derivatives analytically computed using \texttt{Mathematica v12.1}).
All self-consistent field (SCF) calculations were performed using density fitting (DF) in conjunction with the def2-QZVPPD~\cite{weigend2003a,rappoport2010a} and def2-QZVPP-JKFIT~\cite{schuchardtBasisSetExchange2007} basis sets and an energy convergence threshold of \texttt{e\_convergence = 1e-12}.
During DFA training, all calculations employed $(99, 590)$ Lebedev-Treutler grids~\cite{treutlerEfficientMolecularNumerical1995} except for the calculations of the absolute energies in AE18,~\cite{chakravortyGroundstateCorrelationEnergies1993,mardirossianThirtyYearsDensity2017} which used $(500, 974)$.
Minimization of $\mathcal{L}$ in Eq.~\eqref{eq:loss} and optimization of ${\rm wRMSE}(\lambda)$ in Eq.~\eqref{eq:wrmse} were performed in \texttt{Mathematica v12.1}.

\section*{acknowledgments}

All authors thank Richard Kang and Dzmitry (Dima) Vaido for help in assembling the databases and making modifications to the \texttt{Psi4} and \texttt{LibXC} codes.
This material is based upon work supported by the National Science Foundation under Grant No.\ CHE-1945676.
This work was supported in part by the Cornell Center for Materials Research with funding from the Research Experience for Undergraduates program (DMR-1757420 and DMR-1719875).
RAD also gratefully acknowledges financial support from an Alfred P.\ Sloan Research Fellowship.
This research used resources of the National Energy Research Scientific Computing Center, which is supported by the Office of Science of the U.S.\ Department of Energy under Contract No.\ DE-AC02-05CH11231.

\section*{Supporting Information Available:}

\noindent \textit{Supporting Information} (SI) includes: \\

\noindent B-spline definitions; Enforcement of ICF constraints; Training, validation, and testing data sets; Derivation of optimal coefficients and effective degrees of freedom for weighted generalized Tikhonov regularization; Optimized CASE21 ICF coefficients.

\newpage

\section*{TOC Graphic}

\begin{figure}[h!]
    \centering
    \includegraphics[width=\linewidth]{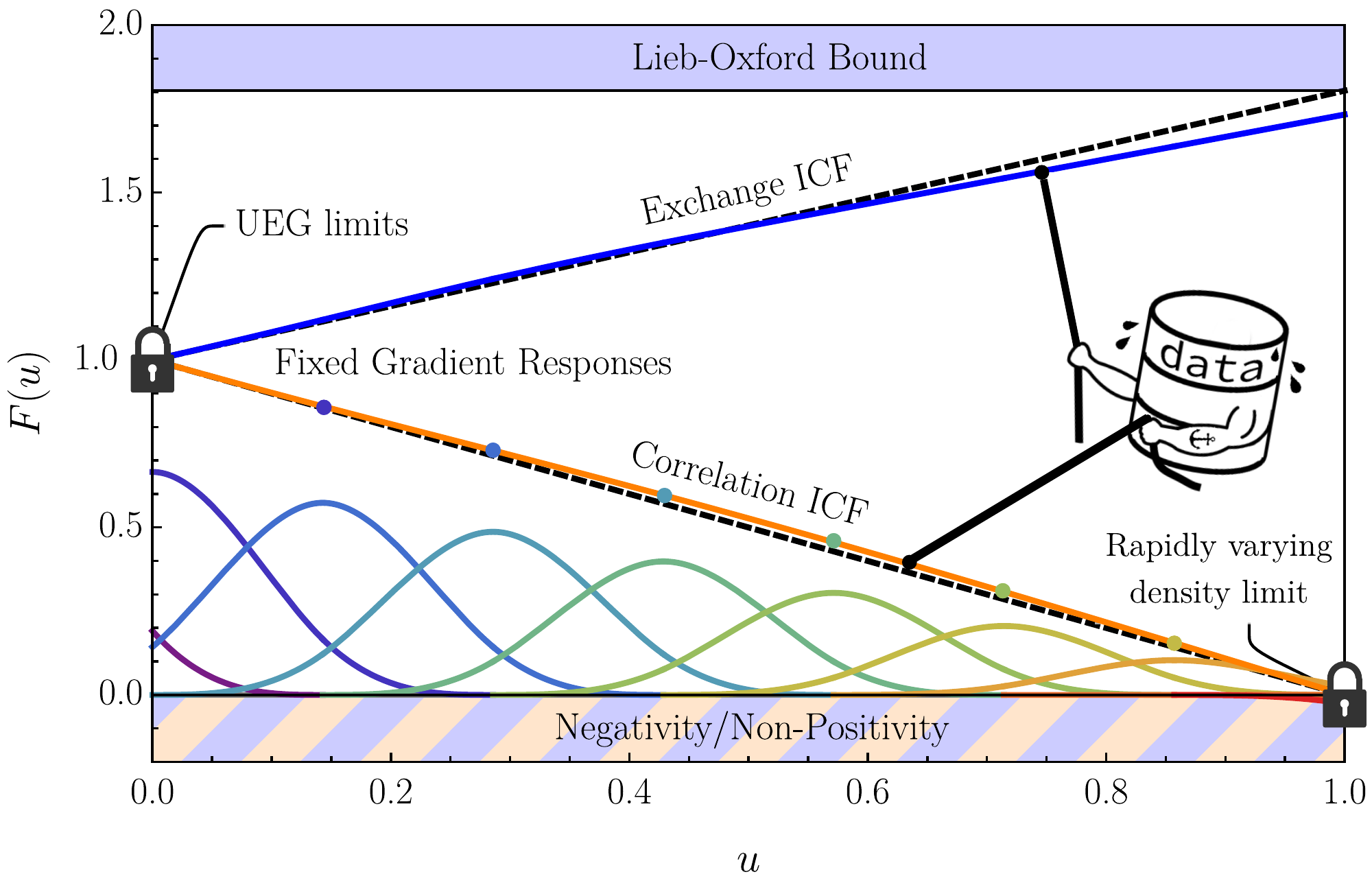}
\end{figure}

\section*{References}


\begin{mcitethebibliography}{89}
\providecommand*\natexlab[1]{#1}
\providecommand*\mciteSetBstSublistMode[1]{}
\providecommand*\mciteSetBstMaxWidthForm[2]{}
\providecommand*\mciteBstWouldAddEndPuncttrue
  {\def\EndOfBibitem{\unskip.}}
\providecommand*\mciteBstWouldAddEndPunctfalse
  {\let\EndOfBibitem\relax}
\providecommand*\mciteSetBstMidEndSepPunct[3]{}
\providecommand*\mciteSetBstSublistLabelBeginEnd[3]{}
\providecommand*\EndOfBibitem{}
\mciteSetBstSublistMode{f}
\mciteSetBstMaxWidthForm{subitem}{(\alph{mcitesubitemcount})}
\mciteSetBstSublistLabelBeginEnd
  {\mcitemaxwidthsubitemform\space}
  {\relax}
  {\relax}

\bibitem[Mardirossian and {Head-Gordon}(2017)Mardirossian, and
  {Head-Gordon}]{mardirossianThirtyYearsDensity2017}
Mardirossian,~N.; {Head-Gordon},~M. Thirty years of density functional theory
  in computational chemistry: {{An}} overview and extensive assessment of 200
  density functionals. \emph{Mol.\ Phys.} \textbf{2017}, \emph{115},
  2315--2372\relax
\mciteBstWouldAddEndPuncttrue
\mciteSetBstMidEndSepPunct{\mcitedefaultmidpunct}
{\mcitedefaultendpunct}{\mcitedefaultseppunct}\relax
\EndOfBibitem
\bibitem[Perdew and Schmidt(2001)Perdew, and
  Schmidt]{perdewJacobLadderDensity2001}
Perdew,~J.~P.; Schmidt,~K. Jacob's ladder of density functional approximations
  for the exchange-correlation energy. \emph{AIP\ Conf.\ Proc.} \textbf{2001},
  \emph{577}, 1--20\relax
\mciteBstWouldAddEndPuncttrue
\mciteSetBstMidEndSepPunct{\mcitedefaultmidpunct}
{\mcitedefaultendpunct}{\mcitedefaultseppunct}\relax
\EndOfBibitem
\bibitem[Becke(1997)]{beckeDensityfunctionalThermochemistrySystematic1997}
Becke,~A.~D. Density-functional thermochemistry. {{V}}. {{Systematic}}
  optimization of exchange-correlation functionals. \emph{J.~Chem.\ Phys.}
  \textbf{1997}, \emph{107}, 8554--8560\relax
\mciteBstWouldAddEndPuncttrue
\mciteSetBstMidEndSepPunct{\mcitedefaultmidpunct}
{\mcitedefaultendpunct}{\mcitedefaultseppunct}\relax
\EndOfBibitem
\bibitem[Perdew \latin{et~al.}(1996)Perdew, Burke, and
  Ernzerhof]{perdewGeneralizedGradientApproximation1996}
Perdew,~J.~P.; Burke,~K.; Ernzerhof,~M. Generalized gradient approximation made
  simple. \emph{Phys.\ Rev.\ Lett.} \textbf{1996}, \emph{77}, 3865--3868\relax
\mciteBstWouldAddEndPuncttrue
\mciteSetBstMidEndSepPunct{\mcitedefaultmidpunct}
{\mcitedefaultendpunct}{\mcitedefaultseppunct}\relax
\EndOfBibitem
\bibitem[Kurth \latin{et~al.}(1999)Kurth, Perdew, and
  Blaha]{kurthMolecularSolidstateTests1999}
Kurth,~S.; Perdew,~J.~P.; Blaha,~P. Molecular and solid-state tests of density
  functional approximations: {{LSD}}, {{GGAs}}, and meta-{{GGAs}}.
  \emph{Int.~J.\ Quantum Chem.} \textbf{1999}, \emph{75}, 889--909\relax
\mciteBstWouldAddEndPuncttrue
\mciteSetBstMidEndSepPunct{\mcitedefaultmidpunct}
{\mcitedefaultendpunct}{\mcitedefaultseppunct}\relax
\EndOfBibitem
\bibitem[Wang and Perdew(1991)Wang, and Perdew]{wangSpinScalingElectrongas1991}
Wang,~Y.; Perdew,~J.~P. Spin scaling of the electron-gas correlation energy in
  the high-density limit. \emph{Phys.\ Rev.~B} \textbf{1991}, \emph{43},
  8911--8916\relax
\mciteBstWouldAddEndPuncttrue
\mciteSetBstMidEndSepPunct{\mcitedefaultmidpunct}
{\mcitedefaultendpunct}{\mcitedefaultseppunct}\relax
\EndOfBibitem
\bibitem[Ma and Brueckner(1968)Ma, and
  Brueckner]{maCorrelationEnergyElectron1968}
Ma,~S.-K.; Brueckner,~K.~A. Correlation energy of an electron gas with a slowly
  varying high density. \emph{Phys.\ Rev.} \textbf{1968}, \emph{165},
  18--31\relax
\mciteBstWouldAddEndPuncttrue
\mciteSetBstMidEndSepPunct{\mcitedefaultmidpunct}
{\mcitedefaultendpunct}{\mcitedefaultseppunct}\relax
\EndOfBibitem
\bibitem[Geldart and Rasolt(1976)Geldart, and
  Rasolt]{geldartExchangeCorrelationEnergy1976}
Geldart,~D. J.~W.; Rasolt,~M. Exchange and correlation energy of an
  inhomogeneous electron gas at metallic densities. \emph{Phys.\ Rev.~B}
  \textbf{1976}, \emph{13}, 1477--1488\relax
\mciteBstWouldAddEndPuncttrue
\mciteSetBstMidEndSepPunct{\mcitedefaultmidpunct}
{\mcitedefaultendpunct}{\mcitedefaultseppunct}\relax
\EndOfBibitem
\bibitem[Langreth and Perdew(1980)Langreth, and
  Perdew]{langrethTheoryNonuniformElectronic1980}
Langreth,~D.~C.; Perdew,~J.~P. Theory of nonuniform electronic systems. {{I}}.
  {{Analysis}} of the gradient approximation and a generalization that works.
  \emph{Phys.\ Rev.~B} \textbf{1980}, \emph{21}, 5469--5493\relax
\mciteBstWouldAddEndPuncttrue
\mciteSetBstMidEndSepPunct{\mcitedefaultmidpunct}
{\mcitedefaultendpunct}{\mcitedefaultseppunct}\relax
\EndOfBibitem
\bibitem[Adamo and Barone(1999)Adamo, and
  Barone]{adamoReliableDensityFunctional1999}
Adamo,~C.; Barone,~V. Toward reliable density functional methods without
  adjustable parameters: {{The PBE0}} model. \emph{J.~Chem.\ Phys.}
  \textbf{1999}, \emph{110}, 6158--6170\relax
\mciteBstWouldAddEndPuncttrue
\mciteSetBstMidEndSepPunct{\mcitedefaultmidpunct}
{\mcitedefaultendpunct}{\mcitedefaultseppunct}\relax
\EndOfBibitem
\bibitem[Sun \latin{et~al.}(2015)Sun, Ruzsinszky, and
  Perdew]{sunStronglyConstrainedAppropriately2015}
Sun,~J.; Ruzsinszky,~A.; Perdew,~J.~P. Strongly constrained and appropriately
  normed semilocal density functional. \emph{Phys.\ Rev.\ Lett.} \textbf{2015},
  \emph{115}, 036402\relax
\mciteBstWouldAddEndPuncttrue
\mciteSetBstMidEndSepPunct{\mcitedefaultmidpunct}
{\mcitedefaultendpunct}{\mcitedefaultseppunct}\relax
\EndOfBibitem
\bibitem[Becke(1993)]{beckeDensityFunctionalThermochemistry1993}
Becke,~A.~D. Density-functional thermochemistry. {{III}}. {{The}} role of exact
  exchange. \emph{J.~Chem.\ Phys.} \textbf{1993}, \emph{98}, 5648--5652\relax
\mciteBstWouldAddEndPuncttrue
\mciteSetBstMidEndSepPunct{\mcitedefaultmidpunct}
{\mcitedefaultendpunct}{\mcitedefaultseppunct}\relax
\EndOfBibitem
\bibitem[Peverati and Truhlar(2014)Peverati, and
  Truhlar]{peveratiQuestUniversalDensity2014}
Peverati,~R.; Truhlar,~D.~G. Quest for a universal density functional: {{The}}
  accuracy of density functionals across a broad spectrum of databases in
  chemistry and physics. \emph{Philos.\ Trans.\ R.\ Soc.~A} \textbf{2014},
  \emph{372}, 20120476\relax
\mciteBstWouldAddEndPuncttrue
\mciteSetBstMidEndSepPunct{\mcitedefaultmidpunct}
{\mcitedefaultendpunct}{\mcitedefaultseppunct}\relax
\EndOfBibitem
\bibitem[Zhao \latin{et~al.}(2005)Zhao, Schultz, and
  Truhlar]{zhaoExchangecorrelationFunctionalBroad2005}
Zhao,~Y.; Schultz,~N.~E.; Truhlar,~D.~G. Exchange-correlation functional with
  broad accuracy for metallic and nonmetallic compounds, kinetics, and
  noncovalent interactions. \emph{J.~Chem.\ Phys.} \textbf{2005}, \emph{123},
  161103\relax
\mciteBstWouldAddEndPuncttrue
\mciteSetBstMidEndSepPunct{\mcitedefaultmidpunct}
{\mcitedefaultendpunct}{\mcitedefaultseppunct}\relax
\EndOfBibitem
\bibitem[Zhao and Truhlar(2008)Zhao, and Truhlar]{zhaoM06SuiteDensity2008}
Zhao,~Y.; Truhlar,~D.~G. The {{M06}} suite of density functionals for main
  group thermochemistry, thermochemical kinetics, noncovalent interactions,
  excited states, and transition elements: {{Two}} new functionals and
  systematic testing of four {{M06}}-class functionals and 12 other
  functionals. \emph{Theor.\ Chem.\ Acc.} \textbf{2008}, \emph{120},
  215--241\relax
\mciteBstWouldAddEndPuncttrue
\mciteSetBstMidEndSepPunct{\mcitedefaultmidpunct}
{\mcitedefaultendpunct}{\mcitedefaultseppunct}\relax
\EndOfBibitem
\bibitem[Mardirossian and {Head-Gordon}(2014)Mardirossian, and
  {Head-Gordon}]{mardirossianOB97XV10parameterRangeseparated2014}
Mardirossian,~N.; {Head-Gordon},~M. {{$\omega$B97X}}-{{V}}: {{A}} 10-parameter,
  range-separated hybrid, generalized gradient approximation density functional
  with nonlocal correlation, designed by a survival-of-the-fittest strategy.
  \emph{Phys.\ Chem.\ Chem.\ Phys.} \textbf{2014}, \emph{16}, 9904--9924\relax
\mciteBstWouldAddEndPuncttrue
\mciteSetBstMidEndSepPunct{\mcitedefaultmidpunct}
{\mcitedefaultendpunct}{\mcitedefaultseppunct}\relax
\EndOfBibitem
\bibitem[Perdew \latin{et~al.}(2005)Perdew, Ruzsinszky, Tao, Staroverov,
  Scuseria, and Csonka]{perdewPrescriptionDesignSelection2005}
Perdew,~J.~P.; Ruzsinszky,~A.; Tao,~J.; Staroverov,~V.~N.; Scuseria,~G.~E.;
  Csonka,~G.~I. Prescription for the design and selection of density functional
  approximations: {{More}} constraint satisfaction with fewer fits.
  \emph{J.~Chem.\ Phys.} \textbf{2005}, \emph{123}, 62201\relax
\mciteBstWouldAddEndPuncttrue
\mciteSetBstMidEndSepPunct{\mcitedefaultmidpunct}
{\mcitedefaultendpunct}{\mcitedefaultseppunct}\relax
\EndOfBibitem
\bibitem[Yu \latin{et~al.}(2016)Yu, Li, and
  Truhlar]{yuPerspectiveKohnShamDensity2016}
Yu,~H.~S.; Li,~S.~L.; Truhlar,~D.~G. Perspective: {{Kohn}}-{{Sham}} density
  functional theory descending a staircase. \emph{J.~Chem.\ Phys.}
  \textbf{2016}, \emph{145}, 130901\relax
\mciteBstWouldAddEndPuncttrue
\mciteSetBstMidEndSepPunct{\mcitedefaultmidpunct}
{\mcitedefaultendpunct}{\mcitedefaultseppunct}\relax
\EndOfBibitem
\bibitem[Medvedev \latin{et~al.}(2017)Medvedev, Bushmarinov, Sun, Perdew, and
  Lyssenko]{medvedevDensityFunctionalTheory2017}
Medvedev,~M.~G.; Bushmarinov,~I.~S.; Sun,~J.; Perdew,~J.~P.; Lyssenko,~K.~A.
  Density functional theory is straying from the path toward the exact
  functional. \emph{Science} \textbf{2017}, \emph{355}, 49--52\relax
\mciteBstWouldAddEndPuncttrue
\mciteSetBstMidEndSepPunct{\mcitedefaultmidpunct}
{\mcitedefaultendpunct}{\mcitedefaultseppunct}\relax
\EndOfBibitem
\bibitem[Dasgupta and Herbert(2017)Dasgupta, and
  Herbert]{dasguptaStandardGridsHighprecision2017}
Dasgupta,~S.; Herbert,~J.~M. Standard grids for high-precision integration of
  modern density functionals: {{SG}}-2 and {{SG}}-3. \emph{J.~Comput.\ Chem.}
  \textbf{2017}, \emph{38}, 869--882\relax
\mciteBstWouldAddEndPuncttrue
\mciteSetBstMidEndSepPunct{\mcitedefaultmidpunct}
{\mcitedefaultendpunct}{\mcitedefaultseppunct}\relax
\EndOfBibitem
\bibitem[Mardirossian and {Head-Gordon}(2016)Mardirossian, and
  {Head-Gordon}]{mardirossianHowAccurateAre2016}
Mardirossian,~N.; {Head-Gordon},~M. How accurate are the {{Minnesota}} density
  functionals for noncovalent interactions, isomerization energies,
  thermochemistry, and barrier heights involving molecules composed of
  main-group elements? \emph{J.~Chem.\ Theory Comput.} \textbf{2016},
  \emph{12}, 4303--4325\relax
\mciteBstWouldAddEndPuncttrue
\mciteSetBstMidEndSepPunct{\mcitedefaultmidpunct}
{\mcitedefaultendpunct}{\mcitedefaultseppunct}\relax
\EndOfBibitem
\bibitem[Wheeler and Houk(2011)Wheeler, and Houk]{wheelerIntegrationGrid2011}
Wheeler,~S.~E.; Houk,~K.~N. Integration grid errors for meta-{{GGA}}-predicted
  reaction energies: {{Origin}} of grid errors for the {{M06}} suite of
  functionals. \emph{J.~Chem.\ Theory Comput.} \textbf{2011}, \emph{6},
  395--404\relax
\mciteBstWouldAddEndPuncttrue
\mciteSetBstMidEndSepPunct{\mcitedefaultmidpunct}
{\mcitedefaultendpunct}{\mcitedefaultseppunct}\relax
\EndOfBibitem
\bibitem[Bootsma and Wheeler(2019)Bootsma, and
  Wheeler]{wheelerIntegrationGrid2019}
Bootsma,~A.~N.; Wheeler,~S.~E. Popular integration grids can result in large
  errors in {{DFT}}-computed free energies. \emph{ChemRxiv} \textbf{2019}, This
  content is a preprint and has not been peer-reviewed.\relax
\mciteBstWouldAddEndPunctfalse
\mciteSetBstMidEndSepPunct{\mcitedefaultmidpunct}
{}{\mcitedefaultseppunct}\relax
\EndOfBibitem
\bibitem[Brown \latin{et~al.}(2021)Brown, Maimaiti, Trepte, Bligaard, and
  Voss]{brownMCMLCombiningPhysical2021}
Brown,~K.; Maimaiti,~Y.; Trepte,~K.; Bligaard,~T.; Voss,~J. {{MCML}}:
  {{Combining}} physical constraints with experimental data for multi-purpose
  meta-generalized gradient approximation. \emph{J.~Comput.\ Chem.}
  \textbf{2021}, \emph{42}, 2004--2013\relax
\mciteBstWouldAddEndPuncttrue
\mciteSetBstMidEndSepPunct{\mcitedefaultmidpunct}
{\mcitedefaultendpunct}{\mcitedefaultseppunct}\relax
\EndOfBibitem
\bibitem[Wellendorff \latin{et~al.}(2012)Wellendorff, Lundgaard,
  M{\o}gelh{\o}j, Petzold, Landis, N{\o}rskov, Bligaard, and
  Jacobsen]{wellendorffDensityFunctionalsSurface2012}
Wellendorff,~J.; Lundgaard,~K.~T.; M{\o}gelh{\o}j,~A.; Petzold,~V.;
  Landis,~D.~D.; N{\o}rskov,~J.~K.; Bligaard,~T.; Jacobsen,~K.~W. Density
  functionals for surface science: {{Exchange}}-correlation model development
  with {{Bayesian}} error estimation. \emph{Phys.\ Rev.~B} \textbf{2012},
  \emph{85}, 235149\relax
\mciteBstWouldAddEndPuncttrue
\mciteSetBstMidEndSepPunct{\mcitedefaultmidpunct}
{\mcitedefaultendpunct}{\mcitedefaultseppunct}\relax
\EndOfBibitem
\bibitem[Wellendorff \latin{et~al.}(2014)Wellendorff, Lundgaard, Jacobsen, and
  Bligaard]{wellendorffMBEEFAccurateSemilocal2014}
Wellendorff,~J.; Lundgaard,~K.~T.; Jacobsen,~K.~W.; Bligaard,~T. {{mBEEF}}:
  {{An}} accurate semi-local {{Bayesian}} error estimation density functional.
  \emph{J.~Chem.\ Phys.} \textbf{2014}, \emph{140}, 144107\relax
\mciteBstWouldAddEndPuncttrue
\mciteSetBstMidEndSepPunct{\mcitedefaultmidpunct}
{\mcitedefaultendpunct}{\mcitedefaultseppunct}\relax
\EndOfBibitem
\bibitem[Lundgaard \latin{et~al.}(2016)Lundgaard, Wellendorff, Voss, Jacobsen,
  and Bligaard]{lundgaardMBEEFvdWRobustFitting2016}
Lundgaard,~K.~T.; Wellendorff,~J.; Voss,~J.; Jacobsen,~K.~W.; Bligaard,~T.
  {{mBEEF}}-{{vdW}}: {{Robust}} fitting of error estimation density
  functionals. \emph{Phys.\ Rev.~B} \textbf{2016}, \emph{93}, 235162\relax
\mciteBstWouldAddEndPuncttrue
\mciteSetBstMidEndSepPunct{\mcitedefaultmidpunct}
{\mcitedefaultendpunct}{\mcitedefaultseppunct}\relax
\EndOfBibitem
\bibitem[Verma and Truhlar(2020)Verma, and
  Truhlar]{vermaStatusChallengesDensity2020}
Verma,~P.; Truhlar,~D.~G. Status and challenges of density functional theory.
  \emph{Trends Chem.} \textbf{2020}, \emph{2}, 302--318\relax
\mciteBstWouldAddEndPuncttrue
\mciteSetBstMidEndSepPunct{\mcitedefaultmidpunct}
{\mcitedefaultendpunct}{\mcitedefaultseppunct}\relax
\EndOfBibitem
\bibitem[Zhao \latin{et~al.}(2006)Zhao, Schultz, and
  Truhlar]{zhaoDesignDensityFunctionals2006}
Zhao,~Y.; Schultz,~N.~E.; Truhlar,~D.~G. Design of density functionals by
  combining the method of constraint satisfaction with parametrization for
  thermochemistry, thermochemical kinetics, and noncovalent interactions.
  \emph{J.~Chem.\ Theory Comput.} \textbf{2006}, \emph{2}, 364--382\relax
\mciteBstWouldAddEndPuncttrue
\mciteSetBstMidEndSepPunct{\mcitedefaultmidpunct}
{\mcitedefaultendpunct}{\mcitedefaultseppunct}\relax
\EndOfBibitem
\bibitem[Eilers and Marx(1996)Eilers, and
  Marx]{eilersFlexibleSmoothingBsplines1996}
Eilers,~P. H.~C.; Marx,~B.~D. Flexible smoothing with {{B}}-splines and
  penalties. \emph{Stat.\ Sci.} \textbf{1996}, \emph{11}, 89--121\relax
\mciteBstWouldAddEndPuncttrue
\mciteSetBstMidEndSepPunct{\mcitedefaultmidpunct}
{\mcitedefaultendpunct}{\mcitedefaultseppunct}\relax
\EndOfBibitem
\bibitem[Eilers \latin{et~al.}(2015)Eilers, Marx, and
  Durb{\'a}n]{eilersTwentyYearsPsplines2015}
Eilers,~P. H.~C.; Marx,~B.~D.; Durb{\'a}n,~M. Twenty years of {{P}}-splines.
  \emph{SORT} \textbf{2015}, \emph{39}, 149--186\relax
\mciteBstWouldAddEndPuncttrue
\mciteSetBstMidEndSepPunct{\mcitedefaultmidpunct}
{\mcitedefaultendpunct}{\mcitedefaultseppunct}\relax
\EndOfBibitem
\bibitem[Ernzerhof and Scuseria(1999)Ernzerhof, and
  Scuseria]{ernzerhofAssessmentPerdewBurke1999}
Ernzerhof,~M.; Scuseria,~G.~E. Assessment of the
  {{Perdew}}\textendash{{Burke}}\textendash{{Ernzerhof}} exchange-correlation
  functional. \emph{J.~Chem.\ Phys.} \textbf{1999}, \emph{110},
  5029--5036\relax
\mciteBstWouldAddEndPuncttrue
\mciteSetBstMidEndSepPunct{\mcitedefaultmidpunct}
{\mcitedefaultendpunct}{\mcitedefaultseppunct}\relax
\EndOfBibitem
\bibitem[Oliver and Perdew(1979)Oliver, and
  Perdew]{oliverSpindensityGradientExpansion1979}
Oliver,~G.~L.; Perdew,~J.~P. Spin-Density Gradient Expansion for the Kinetic
  Energy. \emph{Phys.\ Rev.~A} \textbf{1979}, \emph{20}, 397--403\relax
\mciteBstWouldAddEndPuncttrue
\mciteSetBstMidEndSepPunct{\mcitedefaultmidpunct}
{\mcitedefaultendpunct}{\mcitedefaultseppunct}\relax
\EndOfBibitem
\bibitem[Becke(1986)]{beckeDensityFunctionalCalculations1986}
Becke,~A.~D. Density functional calculations of molecular bond energies.
  \emph{J.~Chem.\ Phys.} \textbf{1986}, \emph{84}, 4524--4529\relax
\mciteBstWouldAddEndPuncttrue
\mciteSetBstMidEndSepPunct{\mcitedefaultmidpunct}
{\mcitedefaultendpunct}{\mcitedefaultseppunct}\relax
\EndOfBibitem
\bibitem[Lieb and Oxford(1981)Lieb, and Oxford]{liebImprovedLowerBound1981}
Lieb,~E.~H.; Oxford,~S. Improved lower bound on the indirect coulomb energy.
  \emph{Int.~J.\ Quantum Chem.} \textbf{1981}, \emph{19}, 427--439\relax
\mciteBstWouldAddEndPuncttrue
\mciteSetBstMidEndSepPunct{\mcitedefaultmidpunct}
{\mcitedefaultendpunct}{\mcitedefaultseppunct}\relax
\EndOfBibitem
\bibitem[Perdew and Wang(1992)Perdew, and
  Wang]{perdewAccurateSimpleAnalytic1992}
Perdew,~J.~P.; Wang,~Y. Accurate and simple analytic representation of the
  electron-gas correlation energy. \emph{Phys.\ Rev.~B} \textbf{1992},
  \emph{45}, 13244--13249\relax
\mciteBstWouldAddEndPuncttrue
\mciteSetBstMidEndSepPunct{\mcitedefaultmidpunct}
{\mcitedefaultendpunct}{\mcitedefaultseppunct}\relax
\EndOfBibitem
\bibitem[Levy(1989)]{levyAsymptoticCoordinateScaling1989}
Levy,~M. Asymptotic coordinate scaling bound for exchange-correlation energy in
  density-functional theory. \emph{Int.~J.\ Quantum Chem.} \textbf{1989},
  \emph{36}, 617--619\relax
\mciteBstWouldAddEndPuncttrue
\mciteSetBstMidEndSepPunct{\mcitedefaultmidpunct}
{\mcitedefaultendpunct}{\mcitedefaultseppunct}\relax
\EndOfBibitem
\bibitem[uni()]{uniscalenote}
Although the CASE21 correlation form is able to exactly cancel the LDA
  logarithmic singularity, the correlation energy completely vanishes in this
  limit. However, to fully satisfy uniform scaling to the high-density limit
  for correlation, the correlation energy should be non-zero in this limit, \eg
  $E_{\rm c} = -0.0467$~Hartree for a two-electron atom as $Z \rightarrow
  \infty$.~\cite{perdewGeneralizedGradientApproximation1996}\relax
\mciteBstWouldAddEndPunctfalse
\mciteSetBstMidEndSepPunct{\mcitedefaultmidpunct}
{\mcitedefaultendpunct}{\mcitedefaultseppunct}\relax
\EndOfBibitem
\bibitem[Prautzsch \latin{et~al.}(2002)Prautzsch, Boehm, and
  Paluszny]{prautzschBezierBSplineTechniques2002}
Prautzsch,~H.; Boehm,~W.; Paluszny,~M. \emph{Bezier and B-Spline Techniques},
  1st ed.; Springer-Verlag, Berlin, Germany, 2002\relax
\mciteBstWouldAddEndPuncttrue
\mciteSetBstMidEndSepPunct{\mcitedefaultmidpunct}
{\mcitedefaultendpunct}{\mcitedefaultseppunct}\relax
\EndOfBibitem
\bibitem[Hansen(1998)]{hansenRankDeficientDiscreteIllPosed1998}
Hansen,~P.~C. \emph{Rank-{{Deficient}} and {{Discrete Ill}}-{{Posed
  Problems}}}; Mathematical {{Modeling}} and {{Computation}}; {SIAM,
  Philadelphia, Pennsylvania}, 1998\relax
\mciteBstWouldAddEndPuncttrue
\mciteSetBstMidEndSepPunct{\mcitedefaultmidpunct}
{\mcitedefaultendpunct}{\mcitedefaultseppunct}\relax
\EndOfBibitem
\bibitem[Strutz(2016)]{strutzDataFittingUncertainty2016}
Strutz,~T. \emph{Data {{Fitting}} and {{Uncertainty}}: {{A}} Practical
  Introduction to Weighted Least Squares and Beyond}, 2nd ed.; {Springer
  Vieweg, Wiesbaden, Germany}, 2016\relax
\mciteBstWouldAddEndPuncttrue
\mciteSetBstMidEndSepPunct{\mcitedefaultmidpunct}
{\mcitedefaultendpunct}{\mcitedefaultseppunct}\relax
\EndOfBibitem
\bibitem[Ernst \latin{et~al.}()Ernst, Sparrow, and {DiStasio
  Jr.}]{NENCI_part_II}
Ernst,~B.~G.; Sparrow,~Z.~M.; {DiStasio Jr.},~R.~A. {{NENCI}}-2021 part {{II}}:
  {{Evaluating}} the performance of quantum chemical approximations on the
  {{NENCI-2021}} benchmark database. (\emph{\textit{in preparation}}). \relax
\mciteBstWouldAddEndPunctfalse
\mciteSetBstMidEndSepPunct{\mcitedefaultmidpunct}
{}{\mcitedefaultseppunct}\relax
\EndOfBibitem
\bibitem[Kang \latin{et~al.}()Kang, Sparrow, Ernst, and {DiStasio
  Jr.}]{NECI_2021}
Kang,~R.; Sparrow,~Z.~M.; Ernst,~B.~G.; {DiStasio Jr.},~R.~A. {{NECI-2021}}:
  {{A}} large benchmark database of non-equilibrium covalent interactions.
  (\emph{\textit{in preparation}}). \relax
\mciteBstWouldAddEndPunctfalse
\mciteSetBstMidEndSepPunct{\mcitedefaultmidpunct}
{}{\mcitedefaultseppunct}\relax
\EndOfBibitem
\bibitem[S.~Yu \latin{et~al.}(2015)S.~Yu, Zhang, Verma, He, and
  G.~Truhlar]{s.yuNonseparableExchangeCorrelation2015}
S.~Yu,~H.; Zhang,~W.; Verma,~P.; He,~X.; G.~Truhlar,~D. Nonseparable
  exchange\textendash correlation functional for molecules, including
  homogeneous catalysis involving transition metals. \emph{Phys.\ Chem.\ Chem.\
  Phys.} \textbf{2015}, \emph{17}, 12146--12160\relax
\mciteBstWouldAddEndPuncttrue
\mciteSetBstMidEndSepPunct{\mcitedefaultmidpunct}
{\mcitedefaultendpunct}{\mcitedefaultseppunct}\relax
\EndOfBibitem
\bibitem[Petzold \latin{et~al.}(2012)Petzold, Bligaard, and
  Jacobsen]{petzoldConstructionNewElectronic2012}
Petzold,~V.; Bligaard,~T.; Jacobsen,~K.~W. Construction of new electronic
  density functionals with error estimation through fitting. \emph{Top.\
  Catal.} \textbf{2012}, \emph{55}, 402--417\relax
\mciteBstWouldAddEndPuncttrue
\mciteSetBstMidEndSepPunct{\mcitedefaultmidpunct}
{\mcitedefaultendpunct}{\mcitedefaultseppunct}\relax
\EndOfBibitem
\bibitem[Levy and Perdew(1985)Levy, and
  Perdew]{levyHellmannFeynmanVirialScaling1985}
Levy,~M.; Perdew,~J.~P. Hellmann-{{Feynman}}, virial, and scaling requisites
  for the exact universal density functionals. {{Shape}} of the correlation
  potential and diamagnetic susceptibility for atoms. \emph{Phys.\ Rev.~A}
  \textbf{1985}, \emph{32}, 2010--2021\relax
\mciteBstWouldAddEndPuncttrue
\mciteSetBstMidEndSepPunct{\mcitedefaultmidpunct}
{\mcitedefaultendpunct}{\mcitedefaultseppunct}\relax
\EndOfBibitem
\bibitem[Bollaerts \latin{et~al.}(2006)Bollaerts, Eilers, and {{van
  Mechelen}}]{bollaertsSimpleMultiplePsplines2006}
Bollaerts,~K.; Eilers,~P. H.~C.; {{van Mechelen}},~I. Simple and multiple
  {{P}}-splines regression with shape constraints. \emph{Br.~J.\ Math.\ Stat.\
  Psychol.} \textbf{2006}, \emph{59}, 451--469\relax
\mciteBstWouldAddEndPuncttrue
\mciteSetBstMidEndSepPunct{\mcitedefaultmidpunct}
{\mcitedefaultendpunct}{\mcitedefaultseppunct}\relax
\EndOfBibitem
\bibitem[Karton \latin{et~al.}(2011)Karton, Daon, and
  Martin]{kartonW411HighconfidenceBenchmark2011}
Karton,~A.; Daon,~S.; Martin,~J. M.~L. W4-11: {{A}} high-confidence benchmark
  dataset for computational thermochemistry derived from first-principles
  {{W4}} data. \emph{Chem.\ Phys.\ Lett.} \textbf{2011}, \emph{510},
  165--178\relax
\mciteBstWouldAddEndPuncttrue
\mciteSetBstMidEndSepPunct{\mcitedefaultmidpunct}
{\mcitedefaultendpunct}{\mcitedefaultseppunct}\relax
\EndOfBibitem
\bibitem[Chakravorty \latin{et~al.}(1993)Chakravorty, Gwaltney, Davidson,
  Parpia, and {Fischer}]{chakravortyGroundstateCorrelationEnergies1993}
Chakravorty,~S.~J.; Gwaltney,~S.~R.; Davidson,~E.~R.; Parpia,~F.~A.;
  {Fischer},~C.~F. Ground-state correlation energies for atomic ions with 3 to
  18 electrons. \emph{Phys.\ Rev.~A} \textbf{1993}, \emph{47}, 3649--3670\relax
\mciteBstWouldAddEndPuncttrue
\mciteSetBstMidEndSepPunct{\mcitedefaultmidpunct}
{\mcitedefaultendpunct}{\mcitedefaultseppunct}\relax
\EndOfBibitem
\bibitem[Karton \latin{et~al.}(2017)Karton, Sylvetsky, and
  Martin]{kartonW417DiverseHighconfidence2017}
Karton,~A.; Sylvetsky,~N.; Martin,~J. M.~L. W4-17: {{A}} diverse and
  high-confidence dataset of atomization energies for benchmarking high-level
  electronic structure methods. \emph{J.~Comput.\ Chem.} \textbf{2017},
  \emph{38}, 2063--2075\relax
\mciteBstWouldAddEndPuncttrue
\mciteSetBstMidEndSepPunct{\mcitedefaultmidpunct}
{\mcitedefaultendpunct}{\mcitedefaultseppunct}\relax
\EndOfBibitem
\bibitem[Morgante and Peverati(2019)Morgante, and
  Peverati]{morganteACCDBCollectionChemistry2019}
Morgante,~P.; Peverati,~R. {{ACCDB}}: {{A}} collection of chemistry databases
  for broad computational purposes. \emph{J.~Comput.\ Chem.} \textbf{2019},
  \emph{40}, 839--848\relax
\mciteBstWouldAddEndPuncttrue
\mciteSetBstMidEndSepPunct{\mcitedefaultmidpunct}
{\mcitedefaultendpunct}{\mcitedefaultseppunct}\relax
\EndOfBibitem
\bibitem[Ye(1998)]{yeMeasuringCorrectingEffects1998}
Ye,~J. On measuring and correcting the effects of data mining and model
  selection. \emph{J.~Am.\ Stat.\ Assoc.} \textbf{1998}, \emph{93},
  120--131\relax
\mciteBstWouldAddEndPuncttrue
\mciteSetBstMidEndSepPunct{\mcitedefaultmidpunct}
{\mcitedefaultendpunct}{\mcitedefaultseppunct}\relax
\EndOfBibitem
\bibitem[Goerigk and Grimme(2010)Goerigk, and
  Grimme]{goerigkGeneralDatabaseMain2010}
Goerigk,~L.; Grimme,~S. A general database for main group thermochemistry,
  kinetics, and noncovalent interactions---assessment of common and
  reparameterized (meta-){{GGA}} density functionals. \emph{J.~Chem.\ Theory
  Comput.} \textbf{2010}, \emph{6}, 107--126\relax
\mciteBstWouldAddEndPuncttrue
\mciteSetBstMidEndSepPunct{\mcitedefaultmidpunct}
{\mcitedefaultendpunct}{\mcitedefaultseppunct}\relax
\EndOfBibitem
\bibitem[Zhao \latin{et~al.}(2005)Zhao, {Gonz{\'a}lez-Garc{\'i}a}, and
  Truhlar]{zhaoBenchmarkDatabaseBarrier2005}
Zhao,~Y.; {Gonz{\'a}lez-Garc{\'i}a},~N.; Truhlar,~D.~G. Benchmark database of
  barrier heights for heavy atom transfer, nucleophilic substitution,
  association, and unimolecular reactions and its use to test theoretical
  methods. \emph{J.~Phys.\ Chem.~A} \textbf{2005}, \emph{109}, 2012--2018\relax
\mciteBstWouldAddEndPuncttrue
\mciteSetBstMidEndSepPunct{\mcitedefaultmidpunct}
{\mcitedefaultendpunct}{\mcitedefaultseppunct}\relax
\EndOfBibitem
\bibitem[Zhao \latin{et~al.}(2004)Zhao, Lynch, and
  Truhlar]{zhaoDevelopmentAssessmentNew2004}
Zhao,~Y.; Lynch,~B.~J.; Truhlar,~D.~G. Development and assessment of a new
  hybrid density functional model for thermochemical kinetics. \emph{J.~Phys.\
  Chem.~A} \textbf{2004}, \emph{108}, 2715--2719\relax
\mciteBstWouldAddEndPuncttrue
\mciteSetBstMidEndSepPunct{\mcitedefaultmidpunct}
{\mcitedefaultendpunct}{\mcitedefaultseppunct}\relax
\EndOfBibitem
\bibitem[Goerigk \latin{et~al.}(2017)Goerigk, Hansen, Bauer, Ehrlich, Najibi,
  and Grimme]{goerigkLookDensityFunctional2017}
Goerigk,~L.; Hansen,~A.; Bauer,~C.; Ehrlich,~S.; Najibi,~A.; Grimme,~S. A look
  at the density functional theory zoo with the advanced {{GMTKN55}} database
  for general main group thermochemistry, kinetics and noncovalent
  interactions. \emph{Phys.\ Chem.\ Chem.\ Phys.} \textbf{2017}, \emph{19},
  32184--32215\relax
\mciteBstWouldAddEndPuncttrue
\mciteSetBstMidEndSepPunct{\mcitedefaultmidpunct}
{\mcitedefaultendpunct}{\mcitedefaultseppunct}\relax
\EndOfBibitem
\bibitem[Curtiss \latin{et~al.}(1997)Curtiss, Raghavachari, Redfern, and
  Pople]{curtissAssessmentGaussian2Density1997}
Curtiss,~L.~A.; Raghavachari,~K.; Redfern,~P.~C.; Pople,~J.~A. Assessment of
  {{Gaussian}}-2 and density functional theories for the computation of
  enthalpies of formation. \emph{J.~Chem.\ Phys.} \textbf{1997}, \emph{106},
  1063--1079\relax
\mciteBstWouldAddEndPuncttrue
\mciteSetBstMidEndSepPunct{\mcitedefaultmidpunct}
{\mcitedefaultendpunct}{\mcitedefaultseppunct}\relax
\EndOfBibitem
\bibitem[Neese \latin{et~al.}(2009)Neese, Schwabe, Kossmann, Schirmer, and
  Grimme]{neeseAssessmentOrbitalOptimizedSpinComponent2009}
Neese,~F.; Schwabe,~T.; Kossmann,~S.; Schirmer,~B.; Grimme,~S. Assessment of
  orbital-optimized, spin-component scaled second-order many-body perturbation
  theory for thermochemistry and kinetics. \emph{J.~Chem.\ Theory Comput.}
  \textbf{2009}, \emph{5}, 3060--3073\relax
\mciteBstWouldAddEndPuncttrue
\mciteSetBstMidEndSepPunct{\mcitedefaultmidpunct}
{\mcitedefaultendpunct}{\mcitedefaultseppunct}\relax
\EndOfBibitem
\bibitem[Karton \latin{et~al.}(2012)Karton, O'Reilly, and
  Radom]{kartonAssessmentTheoreticalProcedures2012}
Karton,~A.; O'Reilly,~R.~J.; Radom,~L. Assessment of theoretical procedures for
  calculating barrier heights for a diverse set of water-catalyzed
  proton-transfer reactions. \emph{J.~Phys.\ Chem.~A} \textbf{2012},
  \emph{116}, 4211--4221\relax
\mciteBstWouldAddEndPuncttrue
\mciteSetBstMidEndSepPunct{\mcitedefaultmidpunct}
{\mcitedefaultendpunct}{\mcitedefaultseppunct}\relax
\EndOfBibitem
\bibitem[Yu and Truhlar(2015)Yu, and Truhlar]{yuComponentsBondEnergy2015}
Yu,~H.; Truhlar,~D.~G. Components of the bond energy in polar diatomic
  molecules, radicals, and ions formed by group-1 and group-2 metal atoms.
  \emph{J.~Chem.\ Theory Comput.} \textbf{2015}, \emph{11}, 2968--2983\relax
\mciteBstWouldAddEndPuncttrue
\mciteSetBstMidEndSepPunct{\mcitedefaultmidpunct}
{\mcitedefaultendpunct}{\mcitedefaultseppunct}\relax
\EndOfBibitem
\bibitem[Parthiban and Martin(2001)Parthiban, and
  Martin]{parthibanAssessmentW1W22001}
Parthiban,~S.; Martin,~J. M.~L. Assessment of {{W1}} and {{W2}} theories for
  the computation of electron affinities, ionization potentials, heats of
  formation, and proton affinities. \emph{J.~Chem.\ Phys.} \textbf{2001},
  \emph{114}, 6014--6029\relax
\mciteBstWouldAddEndPuncttrue
\mciteSetBstMidEndSepPunct{\mcitedefaultmidpunct}
{\mcitedefaultendpunct}{\mcitedefaultseppunct}\relax
\EndOfBibitem
\bibitem[Zhao and Truhlar(2006)Zhao, and
  Truhlar]{zhaoAssessmentDensityFunctionals2006}
Zhao,~Y.; Truhlar,~D.~G. Assessment of density functionals for {$\pi$} Systems:
  {{Energy}} differences between cumulenes and poly-ynes; proton affinities,
  bond length alternation, and torsional potentials of conjugated polyenes; and
  proton affinities of conjugated shiff bases. \emph{J.~Phys.\ Chem.~A}
  \textbf{2006}, \emph{110}, 10478--10486\relax
\mciteBstWouldAddEndPuncttrue
\mciteSetBstMidEndSepPunct{\mcitedefaultmidpunct}
{\mcitedefaultendpunct}{\mcitedefaultseppunct}\relax
\EndOfBibitem
\bibitem[Curtiss \latin{et~al.}(1991)Curtiss, Raghavachari, Trucks, and
  Pople]{curtissGaussianTheoryMolecular1991}
Curtiss,~L.~A.; Raghavachari,~K.; Trucks,~G.~W.; Pople,~J.~A. Gaussian-2 theory
  for molecular energies of first- and second-row compounds. \emph{J.~Chem.\
  Phys.} \textbf{1991}, \emph{94}, 7221--7230\relax
\mciteBstWouldAddEndPuncttrue
\mciteSetBstMidEndSepPunct{\mcitedefaultmidpunct}
{\mcitedefaultendpunct}{\mcitedefaultseppunct}\relax
\EndOfBibitem
\bibitem[Lynch \latin{et~al.}(2003)Lynch, Zhao, and
  Truhlar]{lynchEffectivenessDiffuseBasis2003}
Lynch,~B.~J.; Zhao,~Y.; Truhlar,~D.~G. Effectiveness of diffuse basis functions
  for calculating relative energies by density functional theory.
  \emph{J.~Phys.\ Chem.~A} \textbf{2003}, \emph{107}, 1384--1388\relax
\mciteBstWouldAddEndPuncttrue
\mciteSetBstMidEndSepPunct{\mcitedefaultmidpunct}
{\mcitedefaultendpunct}{\mcitedefaultseppunct}\relax
\EndOfBibitem
\bibitem[Gruzman \latin{et~al.}(2009)Gruzman, Karton, and
  Martin]{gruzmanPerformanceInitioDensity2009}
Gruzman,~D.; Karton,~A.; Martin,~J. M.~L. Performance of ab initio and density
  functional methods for conformational equilibria of
  {{C\textsubscript{n}H\textsubscript{2n+2}}} alkane isomers (n = 4-8).
  \emph{J.~Phys.\ Chem.~A} \textbf{2009}, \emph{113}, 11974--11983\relax
\mciteBstWouldAddEndPuncttrue
\mciteSetBstMidEndSepPunct{\mcitedefaultmidpunct}
{\mcitedefaultendpunct}{\mcitedefaultseppunct}\relax
\EndOfBibitem
\bibitem[Wilke \latin{et~al.}(2009)Wilke, Lind, Schaefer~{{III}},
  Cs{\'a}sz{\'a}r, and Allen]{wilkeConformersGaseousCysteine2009}
Wilke,~J.~J.; Lind,~M.~C.; Schaefer~{{III}},~H.~F.; Cs{\'a}sz{\'a}r,~A.~G.;
  Allen,~W.~D. Conformers of gaseous cysteine. \emph{J.~Chem.\ Theory Comput.}
  \textbf{2009}, \emph{5}, 1511--1523\relax
\mciteBstWouldAddEndPuncttrue
\mciteSetBstMidEndSepPunct{\mcitedefaultmidpunct}
{\mcitedefaultendpunct}{\mcitedefaultseppunct}\relax
\EndOfBibitem
\bibitem[Yu \latin{et~al.}(2015)Yu, Sarrami, Karton, and
  O'Reilly]{yuAssessmentTheoreticalProcedures2015}
Yu,~L.-J.; Sarrami,~F.; Karton,~A.; O'Reilly,~R.~J. An assessment of
  theoretical procedures for {$\pi$}-conjugation stabilisation energies in
  enones. \emph{Mol.\ Phys.} \textbf{2015}, \emph{113}, 1284--1296\relax
\mciteBstWouldAddEndPuncttrue
\mciteSetBstMidEndSepPunct{\mcitedefaultmidpunct}
{\mcitedefaultendpunct}{\mcitedefaultseppunct}\relax
\EndOfBibitem
\bibitem[Lao and Herbert(2015)Lao, and
  Herbert]{laoAccurateEfficientQuantum2015}
Lao,~K.~U.; Herbert,~J.~M. Accurate and efficient quantum chemistry
  calculations for noncovalent interactions in many-body systems: the {{XSAPT}}
  family of methods. \emph{J.~Phys.\ Chem.~A} \textbf{2015}, \emph{119},
  235--252\relax
\mciteBstWouldAddEndPuncttrue
\mciteSetBstMidEndSepPunct{\mcitedefaultmidpunct}
{\mcitedefaultendpunct}{\mcitedefaultseppunct}\relax
\EndOfBibitem
\bibitem[Grimme \latin{et~al.}(2007)Grimme, Steinmetz, and
  Korth]{grimmeHowComputeIsomerization2007}
Grimme,~S.; Steinmetz,~M.; Korth,~M. How to compute isomerization energies of
  organic molecules with quantum chemical methods. \emph{J.~Org.\ Chem.}
  \textbf{2007}, \emph{72}, 2118--2126\relax
\mciteBstWouldAddEndPuncttrue
\mciteSetBstMidEndSepPunct{\mcitedefaultmidpunct}
{\mcitedefaultendpunct}{\mcitedefaultseppunct}\relax
\EndOfBibitem
\bibitem[Luo \latin{et~al.}(2011)Luo, Zhao, and
  Truhlar]{luoValidationElectronicStructure2011}
Luo,~S.; Zhao,~Y.; Truhlar,~D.~G. Validation of electronic structure methods
  for isomerization reactions of large organic molecules. \emph{Phys.\ Chem.\
  Chem.\ Phys.} \textbf{2011}, \emph{13}, 13683--13689\relax
\mciteBstWouldAddEndPuncttrue
\mciteSetBstMidEndSepPunct{\mcitedefaultmidpunct}
{\mcitedefaultendpunct}{\mcitedefaultseppunct}\relax
\EndOfBibitem
\bibitem[Martin(2013)]{martinWhatCanWe2013}
Martin,~J. M.~L. What can we learn about dispersion from the conformer surface
  of n-pentane? \emph{J.~Phys.\ Chem.~A} \textbf{2013}, \emph{117},
  3118--3132\relax
\mciteBstWouldAddEndPuncttrue
\mciteSetBstMidEndSepPunct{\mcitedefaultmidpunct}
{\mcitedefaultendpunct}{\mcitedefaultseppunct}\relax
\EndOfBibitem
\bibitem[Zhao and Truhlar(2006)Zhao, and Truhlar]{zhaoNewLocalDensity2006}
Zhao,~Y.; Truhlar,~D.~G. A new local density functional for main-group
  thermochemistry, transition metal bonding, thermochemical kinetics, and
  noncovalent interactions. \emph{J.~Chem.\ Phys.} \textbf{2006}, \emph{125},
  194101\relax
\mciteBstWouldAddEndPuncttrue
\mciteSetBstMidEndSepPunct{\mcitedefaultmidpunct}
{\mcitedefaultendpunct}{\mcitedefaultseppunct}\relax
\EndOfBibitem
\bibitem[Csonka \latin{et~al.}(2009)Csonka, French, Johnson, and
  Stortz]{csonkaEvaluationDensityFunctionals2009}
Csonka,~G.~I.; French,~A.~D.; Johnson,~G.~P.; Stortz,~C.~A. Evaluation of
  density functionals and basis sets for carbohydrates. \emph{J.~Chem.\ Theory
  Comput.} \textbf{2009}, \emph{5}, 679--692\relax
\mciteBstWouldAddEndPuncttrue
\mciteSetBstMidEndSepPunct{\mcitedefaultmidpunct}
{\mcitedefaultendpunct}{\mcitedefaultseppunct}\relax
\EndOfBibitem
\bibitem[Mardirossian \latin{et~al.}(2013)Mardirossian, Lambrecht, McCaslin,
  Xantheas, and {Head-Gordon}]{mardirossianPerformanceDensityFunctionals2013}
Mardirossian,~N.; Lambrecht,~D.~S.; McCaslin,~L.; Xantheas,~S.~S.;
  {Head-Gordon},~M. The performance of density functionals for
  sulfate\textendash water clusters. \emph{J.~Chem.\ Theory Comput.}
  \textbf{2013}, \emph{9}, 1368--1380\relax
\mciteBstWouldAddEndPuncttrue
\mciteSetBstMidEndSepPunct{\mcitedefaultmidpunct}
{\mcitedefaultendpunct}{\mcitedefaultseppunct}\relax
\EndOfBibitem
\bibitem[Kesharwani \latin{et~al.}(2016)Kesharwani, Karton, and
  Martin]{kesharwaniBenchmarkInitioConformational2016}
Kesharwani,~M.~K.; Karton,~A.; Martin,~J. M.~L. Benchmark ab initio
  conformational energies for the proteinogenic amino acids through explicitly
  correlated methods. {{Assessment}} of density functional methods.
  \emph{J.~Chem.\ Theory Comput.} \textbf{2016}, \emph{12}, 444--454\relax
\mciteBstWouldAddEndPuncttrue
\mciteSetBstMidEndSepPunct{\mcitedefaultmidpunct}
{\mcitedefaultendpunct}{\mcitedefaultseppunct}\relax
\EndOfBibitem
\bibitem[{\v R}ez{\'a}{\v c} and Hobza(2013){\v R}ez{\'a}{\v c}, and
  Hobza]{rezacDescribingNoncovalentInteractions2013}
{\v R}ez{\'a}{\v c},~J.; Hobza,~P. Describing noncovalent interactions beyond
  the common approximations: {{How}} accurate is the ``gold standard,''
  {{CCSD}}({{T}}) at the complete basis set limit? \emph{J.~Chem.\ Theory
  Comput.} \textbf{2013}, \emph{9}, 2151--2155\relax
\mciteBstWouldAddEndPuncttrue
\mciteSetBstMidEndSepPunct{\mcitedefaultmidpunct}
{\mcitedefaultendpunct}{\mcitedefaultseppunct}\relax
\EndOfBibitem
\bibitem[{\v R}ez{\'a}{\v c} and Hobza(2012){\v R}ez{\'a}{\v c}, and
  Hobza]{rezacAdvancedCorrectionsHydrogen2012}
{\v R}ez{\'a}{\v c},~J.; Hobza,~P. Advanced corrections of hydrogen bonding and
  dispersion for semiempirical quantum mechanical methods. \emph{J.~Chem.\
  Theory Comput.} \textbf{2012}, \emph{8}, 141--151\relax
\mciteBstWouldAddEndPuncttrue
\mciteSetBstMidEndSepPunct{\mcitedefaultmidpunct}
{\mcitedefaultendpunct}{\mcitedefaultseppunct}\relax
\EndOfBibitem
\bibitem[Boese(2013)]{boeseAssessmentCoupledCluster2013}
Boese,~A.~D. Assessment of coupled cluster theory and more approximate methods
  for hydrogen bonded systems. \emph{J.~Chem.\ Theory Comput.} \textbf{2013},
  \emph{9}, 4403--4413\relax
\mciteBstWouldAddEndPuncttrue
\mciteSetBstMidEndSepPunct{\mcitedefaultmidpunct}
{\mcitedefaultendpunct}{\mcitedefaultseppunct}\relax
\EndOfBibitem
\bibitem[Boese(2015)]{boeseBasisSetLimit2015}
Boese,~A.~D. Basis set limit coupled-cluster studies of hydrogen-bonded
  systems. \emph{Mol.\ Phys.} \textbf{2015}, \emph{113}, 1618--1629\relax
\mciteBstWouldAddEndPuncttrue
\mciteSetBstMidEndSepPunct{\mcitedefaultmidpunct}
{\mcitedefaultendpunct}{\mcitedefaultseppunct}\relax
\EndOfBibitem
\bibitem[Boese(2015)]{boeseDensityFunctionalTheory2015}
Boese,~A.~D. Density functional theory and hydrogen bonds: {{Are}} we there
  yet? \emph{ChemPhysChem} \textbf{2015}, \emph{16}, 978--985\relax
\mciteBstWouldAddEndPuncttrue
\mciteSetBstMidEndSepPunct{\mcitedefaultmidpunct}
{\mcitedefaultendpunct}{\mcitedefaultseppunct}\relax
\EndOfBibitem
\bibitem[Smith \latin{et~al.}(2014)Smith, Jankowski, Slawik, Witek, and
  Patkowski]{smithBasisSetConvergence2014}
Smith,~D. G.~A.; Jankowski,~P.; Slawik,~M.; Witek,~H.~A.; Patkowski,~K. Basis
  set convergence of the post-{{CCSD}}({{T}}) contribution to noncovalent
  interaction energies. \emph{J.~Chem.\ Theory Comput.} \textbf{2014},
  \emph{10}, 3140--3150\relax
\mciteBstWouldAddEndPuncttrue
\mciteSetBstMidEndSepPunct{\mcitedefaultmidpunct}
{\mcitedefaultendpunct}{\mcitedefaultseppunct}\relax
\EndOfBibitem
\bibitem[Kozuch and Martin(2013)Kozuch, and
  Martin]{kozuchHalogenBondsBenchmarks2013}
Kozuch,~S.; Martin,~J. M.~L. Halogen bonds: {{Benchmarks}} and theoretical
  analysis. \emph{J.~Chem.\ Theory Comput.} \textbf{2013}, \emph{9},
  1918--1931\relax
\mciteBstWouldAddEndPuncttrue
\mciteSetBstMidEndSepPunct{\mcitedefaultmidpunct}
{\mcitedefaultendpunct}{\mcitedefaultseppunct}\relax
\EndOfBibitem
\bibitem[Treutler and Ahlrichs(1995)Treutler, and
  Ahlrichs]{treutlerEfficientMolecularNumerical1995}
Treutler,~O.; Ahlrichs,~R. Efficient molecular numerical integration schemes.
  \emph{J.~Chem.\ Phys.} \textbf{1995}, \emph{102}, 346--354\relax
\mciteBstWouldAddEndPuncttrue
\mciteSetBstMidEndSepPunct{\mcitedefaultmidpunct}
{\mcitedefaultendpunct}{\mcitedefaultseppunct}\relax
\EndOfBibitem
\bibitem[Parrish \latin{et~al.}(2017)Parrish, Burns, Smith, Simmonett,
  DePrince, Hohenstein, Bozkaya, Sokolov, Di~Remigio, Richard, Gonthier, James,
  McAlexander, Kumar, Saitow, Wang, Pritchard, Verma, Schaefer~{{III}},
  Patkowski, King, Valeev, Evangelista, Turney, Crawford, and
  Sherrill]{parrishPsi4OpenSourceElectronic2017}
Parrish,~R.~M.; Burns,~L.~A.; Smith,~D. G.~A.; Simmonett,~A.~C.;
  DePrince,~A.~E.; Hohenstein,~E.~G.; Bozkaya,~U.; Sokolov,~A.~Y.;
  Di~Remigio,~R.; Richard,~R.~M.; Gonthier,~J.~F.; James,~A.~M.;
  McAlexander,~H.~R.; Kumar,~A.; Saitow,~M.; Wang,~X.; Pritchard,~B.~P.;
  Verma,~P.; Schaefer~{{III}},~H.~F.; Patkowski,~K.; King,~R.~A.;
  Valeev,~E.~F.; Evangelista,~F.~A.; Turney,~J.~M.; Crawford,~T.~D.;
  Sherrill,~C.~D. Psi4 1.1: {{An}} open-source electronic structure program
  emphasizing automation, advanced libraries, and interoperability.
  \emph{J.~Chem.\ Theory Comput.} \textbf{2017}, \emph{13}, 3185--3197\relax
\mciteBstWouldAddEndPuncttrue
\mciteSetBstMidEndSepPunct{\mcitedefaultmidpunct}
{\mcitedefaultendpunct}{\mcitedefaultseppunct}\relax
\EndOfBibitem
\bibitem[Lehtola \latin{et~al.}(2018)Lehtola, Steigemann, Oliveira, and
  Marques]{lehtolaRecentDevelopmentsLibxc2018a}
Lehtola,~S.; Steigemann,~C.; Oliveira,~M. J.~T.; Marques,~M. A.~L. Recent
  developments in {{\textsc{libxc}}}\textemdash{}{{A}} comprehensive library of
  functionals for density functional theory. \emph{SoftwareX} \textbf{2018},
  \emph{7}, 1--5\relax
\mciteBstWouldAddEndPuncttrue
\mciteSetBstMidEndSepPunct{\mcitedefaultmidpunct}
{\mcitedefaultendpunct}{\mcitedefaultseppunct}\relax
\EndOfBibitem
\bibitem[Weigend \latin{et~al.}(2003)Weigend, Furche, and
  Ahlrichs]{weigend2003a}
Weigend,~F.; Furche,~F.; Ahlrichs,~R. Gaussian basis sets of quadruple zeta
  valence quality for atoms {{H-Kr}}. \emph{J.~Chem.\ Phys.} \textbf{2003},
  \emph{119}, 12753--12762\relax
\mciteBstWouldAddEndPuncttrue
\mciteSetBstMidEndSepPunct{\mcitedefaultmidpunct}
{\mcitedefaultendpunct}{\mcitedefaultseppunct}\relax
\EndOfBibitem
\bibitem[Rappoport and Furche(2010)Rappoport, and Furche]{rappoport2010a}
Rappoport,~D.; Furche,~F. Property-optimized {{Gaussian}} basis sets for
  molecular response calculations. \emph{J.~Chem.\ Phys.} \textbf{2010},
  \emph{133}, 134105\relax
\mciteBstWouldAddEndPuncttrue
\mciteSetBstMidEndSepPunct{\mcitedefaultmidpunct}
{\mcitedefaultendpunct}{\mcitedefaultseppunct}\relax
\EndOfBibitem
\bibitem[Schuchardt \latin{et~al.}(2007)Schuchardt, Didier, Elsethagen, Sun,
  Gurumoorthi, Chase, Li, and Windus]{schuchardtBasisSetExchange2007}
Schuchardt,~K.~L.; Didier,~B.~T.; Elsethagen,~T.; Sun,~L.; Gurumoorthi,~V.;
  Chase,~J.; Li,~J.; Windus,~T.~L. Basis set exchange: {{A}} community database
  for computational sciences. \emph{J.~Chem.\ Inf.\ Model.} \textbf{2007},
  \emph{47}, 1045--1052\relax
\mciteBstWouldAddEndPuncttrue
\mciteSetBstMidEndSepPunct{\mcitedefaultmidpunct}
{\mcitedefaultendpunct}{\mcitedefaultseppunct}\relax
\EndOfBibitem
\end{mcitethebibliography}

\providecommand{\latin}[1]{#1}
\makeatletter
\providecommand{\doi}
  {\begingroup\let\do\@makeother\dospecials
  \catcode`\{=1 \catcode`\}=2 \doi@aux}
\providecommand{\doi@aux}[1]{\endgroup\texttt{#1}}
\makeatother
\providecommand*\mcitethebibliography{\thebibliography}
\csname @ifundefined\endcsname{endmcitethebibliography}
  {\let\endmcitethebibliography\endthebibliography}{}

\end{document}